# Rich polymorphism of a rod-like liquid crystal (8CB) confined in two types of unidirectional nanopores

R. Guégan[1,2], D. Morineau[1a)], R. Lefort[1], W. Béziel[1,3], M. Guendouz[4], L. Noirez[5], A. Henschel[6] and P. Huber[6]

[1]*Institut de Physique de Rennes, CNRS-UMR 6251, Bâtiment 11A, Université de Rennes 1, F-35042 Rennes, France*

[2] *Institut des Sciences de la Terre, 1A, rue de la Férollerie, F-45071 Orléans, France*

[3] *Laboratoire de Physique des Interfaces, Université Libre de Bruxelles, B-1050 Bruxelles, Belgique.*

[4]*Laboratoire d'Optronique, FOTON, CNRS-UMR 6082, Université de Rennes 1, F-22302 Lannion, France*

[5]*Laboratoire Léon Brillouin (CEA-CNRS), F-91191 Gif-sur-Yvette, France*

[6]*Fakultät für Physik und Elektrotechnik, Universität des Saarlandes, D-66041 Saarbrücken, Germany*

**Abstract**

We present a neutron and X-rays scattering study of the phase transitions of 4-n-octyl-4'-cyanobiphenyl (8CB) confined in unidirectional nanopores of porous alumina and porous silicon (PSi) membranes with an average diameter of 30 nm. Spatial confinement reveals a rich polymorphism, with at least four different low temperature phases in addition to the smectic A phase. The structural study as a function of thermal treatments and conditions of spatial confinement allows us to get insights into the formation of these phases and their relative stability. It gives the first description of the complete phase behavior of 8CB confined in PSi and provides a direct comparison with results obtained in bulk conditions and in similar geometric conditions of confinement but with reduced quenched disorder effects using alumina anopore membranes.

[a] Corresponding author e-mail: denis.morineau@univ-rennes1.fr

# 1 Introduction

Understanding the physical properties of liquid crystals spatially confined at a nanometer scale is of considerable interest both from fundamental and technological points of view [1]. When confined in cavities of a few nanometers, molecular systems experience strong geometrical restrictions and large interfacial interactions [2]. In addition, low dimensionality and quenched disorder effects can occur for some porous materials. They are of special interest when anisotropic phases and continuous transitions are concerned. A large activity has been devoted to freezing-melting behavior of molecular fluids confined in various porous materials [3,4]. In narrow pores, a large fraction of the confined molecules will experience interfacial interactions with the porous matrix [5]. This surface energy term is expected to shift the phase transition temperature, most usually to lower temperature. Strong geometric restrictions may also disrupt the crystalline order, leading to amorphous solid phases [6,7]. Under theses circumstances, even the properties of non-crystalline confined materials (liquids or glassy phases) may differ from the one of the bulk amorphous phases, in terms of density or short range order [8,9]. The situation is more complex for anisotropic molecules. For instance, a close packed organization of long alkanes preferentially aligned parallel to the pore axis of nanoporous Vycor glasses has been reported, while the crystalline lamellar ordering of the molecules is suppressed [10]. Liquid crystals are soft materials, which present anisotropic phases with weak orientational and translational order. They are likely to be considerably influenced by geometric confinement, the nature of the porous surfaces, the pore dimensionality and quenched disorder effects imposed by the pore wall morphology. One of the most rigorously studied liquid-crystals in confined geometry is 4-n-octyl-4’-cyanobiphenyl (8CB), a rod-like molecule. The mesogenic phase behavior of 8CB confined in 200 nm unidimensional nanochannels of alumina anopores membranes has been investigated in details [11,12]. A few degrees depression of the transition temperatures, a broadening of the heat capacity anomalies and a surface induced orientational order near the pore wall have been reported. Random porous materials have also been used in order to address specifically the effects of quenched disorder on the nematic and smectic transitions [13]. Isotropic random porous materials (aerogels) and aerosil dispersions have been used [14,15]. These studies have shown that the nematic to smectic transition is replaced by the gradual occurrence of a short range ordered smectic phase. These results are in apparent agreement with the expectations from random field theories, although a complete conciliation between the different materials and the quantitative theoretical predictions has not been achieved yet [15,16].

At temperatures lower than the smectic transition, the phase behavior of 8CB confined in disordered porous silica glasses shows a rich polymorphism [17,18]. For an average pore size of 20 nm, Fehr et al. have shown that 8CB forms a phase (denoted $K_S$), which is only observable in the bulk state after a strong thermal quench. A new phase $K'_S$, which is apparently never obtained in bulk conditions has also been reported when the pore size is smaller than 10 nm. These two phases are characterized by a unique intense Bragg peak, which has been interpreted by this group as a consequence of a 'frozen' smectic-like arrangement of the molecules imposed by the geometric restriction and particular anchoring of the molecules.

The entanglement of various effects such as finite size, surface interaction, elastic deformation, low dimensionality and quenched disorder, prevents a direct comparison between the used porous materials and a simple interpretation of these results. We have recently shown that the special morphology of porous silicon (PSi) can provide complementary insights with regard to this complex interplay [19]. Indeed, the study of the smectic transition of 8CB confined in PSi has revealed some strong quenched disorder effects, which have been previously reported for homogeneous random porous materials [14, 19].

They are related to the irregular inner pore surface morphology of PSi, which may also affect other transitions such as the capillary condensation of simple gases [20]. Interestingly, PSi is not permeated by randomly oriented, tortuous pores, but by an array of non-interconnected straight nanochannels, which introduce anisotropic random field effects [21]. Because the channels are macroscopically aligned, it offers the opportunity to study the effects of anisotropic confinement without the limitation due to powder average effects [19,22 ]. The confinement conditions provided by PSi present some similarity with strained aerosils, which also possess a unidirectional anisotropy, although they differ by their to pore size and interconnectivity [23,24].

In the present contribution, we focus on the phase transitions of 8CB at temperatures below the bulk smectic transition. We shall give the first description of the complete phase behavior of 8CB confined in PSi and provide a direct comparison with results obtained in bulk conditions and in similar geometric conditions of confinement but with reduced quenched disorder effects using alumina anopore membranes.

We describe, first, the samples and the experimental setups. Then, we present structural and thermodynamical results, obtained by calorimetry, neutron scattering experiments, and X-rays scattering, successively for bulk 8CB, 8CB confined in anopore membranes and 8CB confined in PSi nanochannels. The manuscript ends with a discussion, which addresses the relative effects of finite size, surface morphology, low-dimensionality, quenched disorder for the two confined systems and their consequences in terms of molecular ordering and phase stability.

## 2 Experimental procedure

### 2.1 Sample

4-n-octyl-4'-cyanobiphenyl was purchased from Aldrich and used without further purification. The phase sequence of bulk 8CB on increasing temperature from the crystal phase is crystal (K), smectic A (A), nematic (N) and isotropic (I) with the following transition temperatures: $T_{KA}$=295.2 K, $T_{NA}$=307.0 K and $T_{NI}$=314.0 K [25].

Commercial porous alumina membranes (anopores, with a nominal pore size of 20 nm) were purchased from Whatman company. The nominal pore size claimed by the company corresponds indeed to a thin layer (about 500 nm) present at one side of the membrane, which is used for filtering purpose. Apart from this active surface, the whole bulk of the oxide consists of parallel channels of pore size 150-200 nm running across the membrane of thickness 60 µm [26,27].

Porous silicon (PSi) matrices were made from a (100) oriented silicon substrate using an electrochemical anodization process in a HF electrolyte solution. Anodization conditions control the properties such as the porosity, thickness and the pore size. Anodization of heavily p-doped (100) oriented silicon leads to highly anisotropic pores running perpendicular to the surface wafer (called the columnar form of PSi). These samples were electrochemically etched with a current density of 50 $mA.cm^{-2}$ in a solution composed of HF, $H_2O$ and ethanol (2:1:2) according to previous studies [28]. These controlled conditions give a parallel arrangement of non-interconnected channels (diameter: ~30 nm, length: 30 µm).

Anopore and PSi membranes present a similar porous geometry formed by a parallel arrangement of nanochannels with complementary average diameters. The channels' aspect ratio (diameter/length) in both types of membranes exceeds 1000:1 and confers, therefore, a low dimensionality (quasi 1D) to the system. The preferential alignment of all the channels perpendicularly to the membrane surface prevents powder average limitations when measuring anisotropic observables of unidimensional nanoconfined systems. They are therefore model porous materials for studying nanoconfinement effects in a quasi-1D geometry with macroscopic order along the pore axis.

A specificity of PSi, as compared to anopores, is its strongly irregular inner surface at the nanometer scale. This feature has been recently shown to introduce strong disorder effects, through random surface pinning, which influence phase transitions such as capillary condensation or nematic to smectic transitions [19,20].

8CB was confined into anopores and PSi samples by spontaneous imbibition in the liquid phase. Filling was achieved under 8CB vapor pressure in a vacuum chamber at a temperature of 330 K, above the N-I transition, in order to accelerate the filling of the pores by capillary action. The excess of liquid crystal was removed by squeezing the samples between Whatman filtration papers.

### 2.2 Differential Scanning Calorimetry

The calorimetric experiments were carried out using a DSC 7 from Perkin Elmer. The experiments were performed in the low temperature configuration of the calorimeter using liquid nitrogen as a cooling cryogenic fluid and helium as vector gas. Calibrations of both energy and temperature scales were performed according to standard procedures using the solid-solid and melting transitions of cyclohexane. A systematic procedure was applied for bulk and confined 8CB in order to produce the same thermal treatment. The sample was first heated to 320 K, in the liquid phase. The calorimetric scan was monitored on decreasing temperature down to 200 K. The sample was held at this temperature for 5 minutes and heated again to 320 K. Two different rates have been used, respectively 2 and 10 $K.min^{-1}$, in order to quantify kinetic effects. Specific thermal treatments and ageing have also been investigated for 8CB confined in anopore in order to address metastability issues.

### 2.3 Small Angle Neutron Scattering

Small angle neutron scattering experiments were carried out on the spectrometer PAXY of the Léon Brillouin Laboratory (CEA-CNRS, Saclay, France). A displex cryostat was used to vary the temperature of the samples in a range from 180 to 340 K. A monochromatic incident wavelength of 5 Å was selected and the bi-dimensional multidetector was located at 1.5 meters from the sample. These experimental conditions allow one to cover a range of momentum transfers $q$ from 0.03 to 0.3 $Å^{-1}$.

### 2.4 Neutron Diffraction

Neutron diffraction experiments were performed on the double axis spectrometer G6.1 of the Léon Brillouin Laboratory neutron source facility (CEA-CNRS, Saclay, France) using a monochromatic incident wavelength of 4.74 Å in order to cover a $q$-range from 0.1 to 1.8 $Å^{-1}$. A cryostat was used to heat and cool the samples. The temperature was controlled with a stability better than 0.1 K. The measurement temperatures were ranged from 180 to 315 K, with an average scanning rate of about $0.1 K.min^{-1}$. This spectrometer is equipped with a linear multidetector, which does not provide direct information on the structural anisotropy of the sample. However, the porous membranes (typically eight) were stacked vertically in a cylindrical aluminium cell, which has allowed us to investigate different crystallographic directions by changing the angle of incidence by a simple rotation of the cell.

### 2.5 X-ray diffraction

The sample was mounted on a copper frame in the sample cell. The cell consisted of a Peltier cooled base plate and a Be cap. It was filled with He gas for thermal contact. The cell was in a vacuum chamber, the outer jacket of which has Mylar windows allowing the passage of the x-rays over a wide range of scattering angles within the scattering plane. The lowest temperature reachable with this setup was $T$ = 245 K.

The diffraction experiments were carried out on a two-circle diffractometer with graphite monochromatized Cu $K$ radiation emanating from a rotating anode. The two angles that can be varied are the detector angle 2 and the rotation angle $\omega$ about the normal of the scattering plane. Typical temperature scanning rates were in the range of 0.01K.min$^{-1}$.

## 3 Results

The calorimetric measurements performed at 2K.min$^{-1}$ for bulk 8CB are displayed in Fig. 1 as reference. Two exothermic peaks are observed on cooling, which correspond to the isotropic-nematic and smectic-crystal transitions respectively. During the subsequent heating scan, three calorimetric signatures are observed: the K-A transition is characterized by a strong endothermic peak at 294.6 K, the second order A-N transition is detected by a very weak jump of the specific heat at 306.5 K and the A-N transition is characterized by a last endothermic peak at 313.7 K [25,29]. A large thermal hysteresis between heating and cooling is observed for the K-A transition, which is a signature of the first order character of the transition. The phase transition temperatures can be affected by the ramp rates. The activated dynamics of the transitions of bulk 8CB has been recently investigated by differential scanning calorimetry and modulation calorimetry [29]. In order to quantify rate effects and to disentangle kinetic effects from genuine thermal hysteresis, two different ramp rates have been used and the transition temperatures have been extrapolated to zero ramp rate (cf. Figs 2). The obtained values are in agreement with accepted values for bulk 8CB (cf. Table 1) and will allow a direct comparison with scattering experiments performed at very small rates.

The small angle neutron scattering patterns on the 2D detector of bulk 8CB are displayed in Fig. 3. At 315 K, no diffraction peak is observed but a flat background, which reflects the lack of long-range translational order in the liquid phase (cf. Fig. 3 (a)). At 300 K, a clear diffraction ring appears for an angle of diffraction corresponding to a value of the momentum transfer $q$ equals to 0.2 $Å^{-1}$, as shown in Fig. 3(b). This diffraction pattern is the signature of the quasi-long range translational order in the smectic phase. The $q$-value corresponds to a modulation of the density across the smectic layers of periodicity about 3 nm. The formation of a large number of smectic domains with no preferential orientations is illustrated by the circular symmetry of the pattern. The orientation of the smectic domains has been performed by decreasing the temperature from the isotropic phase down to the smectic phase under an external magnetic field of 0.6 T, as shown in Fig. 3(c).

The structural signatures of the different phases observed for bulk 8CB during heating from 255 to 315 K are provided by the powder diffraction experiments performed on G6.1, as shown in Fig. 4. At 255 K, six Bragg peaks are observed at 0.48, 1.13, 1.34, 1.41, 1.63 and 1.69 $Å^{-1}$ and can be assigned to the stable crystalline phase of 8CB, denoted phase K [18]. The intensity of these peaks decreases slightly at 294 K while a small smectic peak rises at 0.2 $Å^{-1}$, as a consequence of the melting of the crystal. The single peak characteristic of the bulk smectic phase is recovered at 300 K, whereas only broad peaks are observed at 310 and 315 K corresponding to short-range smectic critical fluctuations in the nematic and isotropic phases respectively.

DSC scans acquired in the same experimental conditions for 8CB confined in anopores are presented in Fig. 5. On cooling at 10K.min$^{-1}$, a small exothermic peak related to the I-N transition is observed first and a broad intense exothermic peak is observed around 253 K. This second peak is attributed to a crystallization process, which occurs about 10 K below the bulk one. A lowering of the temperature of crystallization is usually encountered in confined geometry, because of changes in the nucleation processes and/or shifted phase diagrams. In fact, the scan obtained on heating qualitatively differs from the bulk one (see solid line with striped area in Fig. 5), with a large endothermic peak at 286.1 K and a splitted peak around 291.7 K (denoted peaks A and B in Fig. 5) in addition to a third small endothermic peak at 311.7 K. The last peak is obviously the signature of the N-I transition, which is slightly depressed in confinement ( $T \approx -2K$). The two large endothermic peaks reveal the existence of at least one additional solid state phase transition in confinement, missing in the bulk. In order to check the relative stability of these phases, a subsequent scan on heating was interrupted at 298K, where the sample was aged for 5 minutes (between peaks A and B). It was then cooled down to 200K and the complete scan was acquired on heating again (as displayed as a solid line with gray-shaded area in Fig. 5). Owing to this thermal treatment, peak A totally disappears, whereas the peak B intensity is increased by a factor of 2. This means that, on first cooling, different solid phases are formed and co-exist within the porous alumina anopore. Moreover, it implies that the one, which melts at 286.1 K (peak A) is actually metastable towards the phases that melt at higher temperature (peak B).

Rates effects have been investigated by performing the same thermal treatments with a ramp of $2K.min^{-1}$ (cf. inset in Fig. 5). Transition temperature shifts have been quantified and the transition temperature extrapolated to zero ramp rate (cf. Figs. 2). In addition, for a rate of $2K.min^{-1}$ an exothermic peak is visible a few degrees above the melting peak (A) (see solid line with striped area in inset), which obviously corroborate the conclusions about the relative stability of the solid phases deduced from isothermal ageing experiments (see solid line with gray-shaded area in inset).

The assignment of these low temperature solid phases is possible from the neutron diffraction results, which are discussed later. The high temperature part of the phase diagram that concerns the mesomorphic phases is not qualitatively different in confinement from the bulk one. The phase sequence is maintained but the phase transitions are slightly broadened and shifted to lower temperature.

The small angle neutron scattering pattern of 8CB confined in anopores at 300 K (cf. Fig. 6) reveals a well-defined Bragg peak at 0.2$Å^{-1}$. This sharp peak shows that the bulk smectic order is maintained in confinement. In addition, it proves that the orientation of the smectic domains is not isotropically distributed as it is for the bulk system but follows precisely the main axis of the nanochannels. This preferential unidirectional orientation induced by anisotropic confinement is extremely well-defined and comparable with the bulk one under the application of a magnetic field (cf. Fig. 3(c)).

Neutron diffraction experiments on G6.1 have allowed us to cover a larger *q*-range and to identify the low-temperature phases observed by DSC for 8CB in anopores. Fig. 7(a) shows the neutron scattered intensity recorded in normal incidence, which means that the momentum transfer is perpendicular to the pore axis in the limit of low *q* values. The displayed spectra have been measured during heating process from 255 to 315 K. The spectra obtained from 255 to 291.7 K display an intense Bragg peak at 0.48$Å^{-1}$ and two small peaks at 1.34 and 1.41$Å^{-1}$, which are some signatures of the bulk crystalline phase K. Their intensities decrease above 292 K. This feature can be associated to the second endothermic peak observed by DSC and corresponds to the transition from the crystal K to the smectic A phase, which is depressed in confinement ( $T \approx -2$ K). No smectic Bragg peak is observed at 300 K in this incidence, as a direct consequence of the extreme alignment of the smectic phase confined in anopores.

This is at variance with the diffraction pattern acquired in grazing incidence (cf. Fig. 7(b)), which presents a very intense peak at 300 K for a momentum transfer of 0.2 Å$^{-1}$. It corresponds to the scattering from the aligned smectic layers, which fulfill the Bragg reflection conditions in grazing incidence only.

Under grazing incidence, other much weaker peaks are observed at lower temperatures (cf. inset). Their small intensity suggests that the corresponding crystallites, which are probably aligned within the channels, do not totally fulfill the Bragg reflection condition under this incidence. At the lowest temperature 255 K, two small Bragg peaks are observed at 0.25 and 0.65 Å$^{-1}$. On increasing temperature, these two peaks vanish independently at temperatures around 287 K and 293.5 K respectively. These two transition temperatures are consistent with the two endothermic peaks observed by DSC (cf. Fig. 5). In order to summarize this phase sequence, the temperature variation of the intensity of some characteristic Bragg peaks is reported in Fig. 8 for both bulk 8CB and 8CB in anopore.

The occurrence of a single peak at 0.25Å$^{-1}$ has already been observed in confinement in aerogels [17]. It has been assigned to a metastable phase of 8CB, which is promoted by confinement and denoted Ks.

The second small peak at 0.65Å$^{-1}$ has not been observed in the diffraction pattern of the bulk crystalline K phase. However, it presents the same temperature variation as the peaks assigned to the K phase in anopores, which are observed in grazing incidence and melts at 293.5 K, leading to the endothermic peak (B) in Fig. 4. Fehr et al. have mentioned that 8CB could crystallize in a variant form of the K phase, denoted K', which presents such a Bragg peak at 0.65Å$^{-1}$ [17,18].This is also consistent with the splitting of the endothermic peak (B) in Fig. 4, which indicate the almost concomitant melting of phases K and K'.

Small angle neutron scattering patterns of 8CB confined in PSi are displayed in Fig. 9. These measurements have been performed on cooling from 340 to 180 K. Spectra obtained above 290 K do not show any significant scattering from awaited short range translational correlations within the liquid and nematic phases. Two symmetrically located Bragg peaks appear at 290 K at $q$=0.2 Å$^{-1}$, which is the signature of a smectic order within the confined phase. This phase corresponds actually to a short-range smectic phase (SRS), which has been discussed previously [19]. The two peaks have a crescent shape with an angular aperture (half width at half maximum) of 30 degrees. It proves the existence of a preferential orientation of the smectic director along the nanochannels axis, although not as pronounced as for 8CB confined in anopores (cf. Fig. 6). On further decreasing the temperature, four symmetric Bragg peaks appear at 240 K. They correspond to a preferential crystallographic direction, which is tilted by about 45 degrees from the nanochannels axis. The associated momentum transfer is $q$=0.25 Å$^{-1}$ and can be assigned to the Ks phase already observed in anopores. The Ks phase coexists with the SRS phase at 240 K. At 220K, the Bragg peak attributed to the remaining SRS phase vanishes and is replaced by two symmetrically crescent peaks located on the vertical axis at 0.15 Å$^{-1}$. This additional phase, has never been observed in bulk nor in anopores, but could possibly be related to the so-called K's phase first reported recently for 8CB confined in porous silica glasses with a pore size diameter of 10 nm. The angular distribution of the orientation of this phase around the pore axis is the same as the SRS phase, i.e. 30 degrees. This phase coexists with the Ks phase at 220 K and no change in the neutron scattering intensity is observed on further decreasing the temperature down to 180 K.

A comprehensive analysis of the variation with the temperature of the integrated intensity of characteristic Bragg peaks of 8CB confined in PSi have been performed on G6.1. Fig. 10 shows the intensity of the scattered intensity at $q$=0.2, 0.15 and 0.25 Å$^{-1}$ corresponding to Bragg peaks of the smectic A, K's and Ks phases respectively. The data shown in Fig. 10 have been collected for 8CB confined in PSi on cooling. The smectic peak intensity of bulk 8CB is added for comparison. The phase sequence is in agreement with the results of small angle neutron scattering previously discussed. Additional information are provided by the experiments performed on G6.1 thanks to the large number of selected temperatures. The N-A transition is absent in PSi and replaced by a progressive increase of smectic correlation. This striking behavior has been interpreted as a consequence of quenched disorder effects induced by confinement in PSi and is fully discussed elsewhere [19]. A quantitative analysis of the shape of the smectic Bragg peak has shown that the correlation length progressively grows from 3 to 12 nm, i.e. from one to four smectic layer thicknesses. Crystallization is shifted to extremely low temperature as compared to bulk 8CB. The temperature range of stability of the SRS phase in PSi is therefore extremely broad, more than five times larger than the bulk smectic one. Crystallization firstly occurs on cooling through the occurrence of a unique peak at 0.25 $Å^{-1}$ at 250 K, which is attributed to the Ks phase. In the meantime, the intensity of the SRS peak decreases by about 15% of its maximum value. This allows one to quantify the fraction of 8CB converted from the SRS to the Ks phase. On further cooling, SRS and Ks phases coexist down to 232 K, where the remaining 85% of SRS phase finally crystallizes into the K's phase. This original phase is apparently characterized by a unique peak located at 0.15 $Å^{-1}$.

The phase behavior observed during the subsequent heating of 8CB in PSi is shown in Fig. 11. For the sake of clarity, the SRS, Ks and K's phases are considered on separated figures. The intensity of the characteristic Bragg peaks is plotted during a cooling-heating cycle, in addition with the bulk smectic or crystal phase during cooling for comparison. On heating, the intensity of the SRS phase in Fig. 11(a) firstly increases along with the decrease of the one of the K's Bragg peak in Fig. 11(b). It proves that the K's phase, i.e. about 85% of the confined system, melts around 255 K and transforms into the SRS phase. Note that both crystallization and melting processes are broadened, which is a usual observation in a confined system. Therefore, the transitions temperatures have been defined at the mid-point of the complete phase transformation. The heating scan gives 255 K as the temperature of thermodynamical equilibrium between the two unusual K's and SRS phases for 8CB confined in PSi. The large thermal hysteresis is an indication of the first order character of the transition and allows one to maintain the SRS in a metastable state down to 240K experimentally.

On further heating, the intensity of the Ks Bragg peak vanishes at about 285 K, while another Bragg peak raises at 0.48Å$^{-1}$, as shown in Fig. 11(c). This proves that the remaining 15% of the system in the Ks phase does not melt into the SRS phase but first transforms into the K phase (i.e. the usual crystalline form of bulk 8CB). The confined K phase ultimately melts around 296 K into the SRS phase.

Complementary experiments have been performed by X-ray diffraction on 8CB confined in PSi. The same samples have been used in order to guarantee the control of the conditions of spatial confinement. The diffraction patterns have been firstly acquired during cooling at very slow rate (c.a. 0.01 K.min$^{-1}$) from the isotropic phase down to 245 K followed by a subsequent heating up to 310 K. Three patterns are shown in Fig. 11 at 280K and 245K on cooling, and at 280K after re-heating. The first pattern at 280 K presents a unique Bragg peak at $q$=0.2Å$^{-1}$, which is typical of the SRS phase and in agreement with neutron scattering experiments. At 245K, the peak at $q$=0.15Å$^{-1}$ observed by neutron scattering is absent in the x-ray data, whereas the peak at $q$=0.25Å$^{-1}$ is in fact splitted in two peaks located respectively at $q$=0.24Å$^{-1}$ and $q$=0.26Å$^{-1}$. This splitting was also insinuated by neutron scattering experiments, but could be easily overlooked because of a poorer signal and $q$-resolution. Moreover a moderately intense peak appears at $q$=1.63Å$^{-1}$ in addition to a collection of much weaker peaks. The more intense peak at $q$=1.63Å$^{-1}$ and some of the weaker peaks are compatible with the assignment of the K' phase, as shown in Fig. 12 [18]. On heating, the whole Bragg peaks essentially vanish, along with the growth of a series of Bragg peaks at 0.48, 0.96, 1.15, 1.26, 1.35, 1.42 and 1.69 Å$^{-1}$, which have been attributed to the crystalline K phase. The temperature variation of the intensity of these different Bragg peaks reveals a unique smectic-crystal transition on cooling around 272K. On heating, solid-solid transitions are smeared out from 266K to 270K followed by melting of the K phase at 295K.

## 4 Discussion

Confinement of 8CB in porous materials induces several effects, which strongly affect the phase structure and phase transition. Among the different issues that can be addressed, pore dimensionality is fundamental to anopores and PSi. The use of porous materials, which display macroscopically long and perfectly aligned unidirectional channels, is essential in order to prevent powder averaging effects and study the possible alignment of the LC under confinement. In the case of 8CB in anopores, an extreme alignment of the smectic phase is observed, the angular distribution of the smectic order direction being smaller than 10 degrees. In the case of 8CB in PSi, a significant angle distribution of the smectic Bragg peak around the pore channel axis is observed. The mosaïcity is about 30 degrees, which is too large to be attributed to the angular distribution of the channel axis, which are straight and parallel to each other up to the macroscopic scale. The alignment results from a combination of the dimensionality of the porous volume and the anchoring conditions induced at the solid interface [30,31,32]. Most probably, the case of PSi does not reflect fluctuations of the uniaxial character of the porous topology but rather some pore surface irregularities. The roughness of the pore surface of PSi leads to varying anchoring conditions, and quenched disorder effects, which promote the formation of short ranged correlated smectic domains with a distribution of preferential orientations [19].

The high temperature phase behaviour of 8CB confined in anopores with a diameter of 200 nm has been previously studied by calorimetry [11,12]. No change in the sequence of mesomorphic phases has been reported, but a reduction of the transition enthalpy and temperature as well as a broadening of the transitions. It has revealed the large effects of elastic distortion and the nature of the surface order on the phase transitions. We have shown that confinement in anopore channels also lead to the occurrence of a different phase sequence at low temperature, with two coexisting crystalline polymorphic phases. The additional solid phase, characterized by a Bragg peak at $q$=0.25 Å$^{-1}$, has been already observed for 8CB confined in aerogels, a different type of porous materials, which mostly corresponds to a network of interconnected cavities [17]. The same group has reported the possibility to obtain the Ks phase in bulk condition when appropriate thermal treatment and quenching are applied. This suggests that the Ks phase corresponds to a metastable polymorphic form of bulk 8CB, the formation of which is promoted by confinement. Note that the metastability of Ks towards the more usual K and K' solids phases is proved for 8CB confined in alumina by DSC (cf. Fig. 5). This behavior compares with the one reported for other systems, such as water, which is known to preferentially crystallize in cubic ice under strong confinement conditions rather than the bulk most stable hexagonal phase [33]. The structure of this Ks solid form of 8CB has not been resolved yet. Diffraction patterns of Ks in anopores have only revealed a unique intense Bragg peak, located at a value of momentum transfer ($q$=0.25Å$^{-1}$) close to the regular smectic A phase ($q$=0.2Å$^{-1}$). This suggests that the Ks phase presents indeed a smectic-like lamellar unidirectional structure, with a slightly smaller interspacing distance. The apparent absence of any other diffraction peak in the experimental results, confirms the results obtained in aerogels although it does not allow one to conclude about the absence of an underlying three-dimensional translational order.

A very different behaviour is obtained for 8CB confined in PSi. Some most remarkable features shown in Fig. 10 totally remodel the phase sequence. They correspond to the disappearance of the smectic phase transition and the occurrence of new low-temperature phases. A short-range smectic order progressively appears on decreasing temperature, with a correlation length saturating to a value of only four smectic layer thicknesses, in place of a true nematic-smectic transition with a developing correlation length. This behaviour has been related to strong quenched disorder effects [19]. This interpretation has been supported by the lineshape analysis of the broad smectic Bragg peak, which presents two components, the first being related to normal smectic thermal fluctuations and the second to the static disorder term. Expressions derived from random fields theories have shown to provide a quantitative fit of the temperature variation of the neutron scattering intensity. Quenched disorder effects obviously dominate the structure and phase behaviour of 8CB in PSi from room temperature down to 250K. PSi has provided the first evidence of strong unidirectional quenched disorder effects for a LC confined in a porous solid. They originate from the highly irregular inner surface, which leads to an almost random pinning of the interfacial molecules that couple to the LC order parameters. The main striking consequences are the suppression of the N-A transition, in agreement with theoretical expectations, and a gradual increase on cooling of the short-ranged smectic correlation length, which is preferentially aligned along the pore axis. The temperature variation of smectic correlations in the SRS phase arises from the competition between the strength of disorder and the smectic elasticity. These effects could also play a role in the recently reported extreme slowdown of the molecular dynamics [34,35,36].

A second specificity of the phase diagram of 8CB in PSi is the strong inhibition of the crystallization processes, which finally occurs on further cooling below 250 K. Two unusual polymorphic phases are encountered, denoted Ks and K's. These phases are essentially characterized by Bragg peaks, located around $q$=0.25 Å$^{-1}$ and $q$=0.15 Å$^{-1}$, respectively. The Ks phase has been reported in previous studies for 8CB confined in aerogels or for quenched bulk 8CB. The observation of the K's phase is even more peculiar than the Ks, since it never occurs in bulk, nor in anopore and have been first reported by Fehr et al. in aerogels only for the smallest pore size (10nm) and promoted by thermal quenching [17,18]. The K's phase occurs in PSi, despite a three times larger pore size, on a moderate cooling rate.

Ks and K's phases are metastable with respect to the K phase for 8CB in PSi, as it has been already proved for the Ks phase of 8CB in anopore. This is proved by the melting of the K's phase at 251 K and the irreversible transformation of the Ks phase to the stable crystal K phase on heating. The relative stability of these metastable phases is also consistent with less ordered structures. Metastable phases are always formed through a competition with the nucleation and growth of more stable phases. This usually gives a central role to the thermal rate on the nature of the obtained phase or the proportion of coexisting phases. Indeed, the Ks phase has been obtained in bulk conditions only after a thermal quench, whereas slower cooling rates promotes the more stable K phase [18]. For 8CB in PSi, temperature scan rates effects are also highlighted by the comparison between X-rays and neutron diffraction experiments. A coexistence of K's and Ks phases in proportion 85:15 is achieved after cooling from room temperature to 255K with average rate of 0.1K.min$^{-1}$. The much slower cooling rate used for X-rays scattering experiments (about 0.01 K.min$^{-1}$) obviously promotes the Ks phase, which is more stable than the K's phase. In addition, the SmA-Ks phase transition occurs on cooling at a higher temperature for X-rays than for neutron scattering experiments, which unambiguously reflects a kinetically controlled nucleation-growth process instead of a genuine supercooling of the smectic phase.

There is an apparent analogy between thermal quench and confinement effects, which are both expected to favour metastable phases. The rapid increase of viscosity on cooling or under strong confinement conditions limits the relaxation of molecular rearrangements, leading to frozen-in structures. As pointed out by Fehr et al., the similarity between thermal quench and confinement effects is amplified by the presence of quenched disorder effects in irregular porous materials, which bring an additional constraint to full structural relaxation and extension of a completely ordered structure [18]. The very different phase behaviour of 8CB in anopores and PSi, concerning both the SmA and the solid phases, reflects the previously mentioned different strengths of confinement and disorder introduced by these mesoporous materials [19].

Table 1 summarizes the different phase sequences that have been observed in bulk, PSi and anopores. The complexity of the phase behaviour is due to the existence of at least four polymorphic phases with temperature dependent nucleation and growth rates. First important information is the melting temperate measured on heating, which relates the stability of the different solid phases with respect to the smectic one and are intrinsically thermodynamic quantities. The melting of the K’s phase has been measured in PSi only (around 251 K), and the melting of the Ks and K’ phases has been observed in anopores only (around 287 K and 293 K respectively). The melting of the K phase is around 295 K for the two confined systems. Combining the melting temperatures measured in anopores and PSi, one can draw a schematic Gibbs diagram of nanoconfined 8CB, which sorts the relative stability of the different confined phases in the following order: K’s<<Ks<K’<K (as schematically shown in Fig. 13).

Confining 8CB in different porous materials primarily alters the nucleation and growth processes, which may promote different metastable phases. The K’s phase of highest energy can be reached only for strong confinement, as also noticed in aerogels, and for sufficiently fast cooling rates (experiments performed by neutron scattering). Slower cooling rates or more regular porous geometry (anopores versus PSi), allow the growth of the more stable Ks lamellar phase, which may co-exist with the K’s in different proportion depending on the thermal treatment. The more stable three-dimensional crystalline phases (K and K’) can form on cooling with a very slow rate (X-rays) or, as most usually observed for PSi, nucleates at low temperature in the Ks phase and grows on heating in a region around 270K to 280 K.

This Ks-K transition on heating occurs on heating at a lower temperature for X-rays than for neutron scattering experiments, which again clearly illustrates the sensitivity to different heating rates of this kinetically controlled growth of the K phase driving the complete phase transition.

We shall now discuss the structure of the low temperature phases of 8CB confined in PSi. The prevalence of Bragg peaks at low *q* values for the K's and Ks phases has been logically associated to a lamellar structure. Constraints induced by confinement, in terms of interfacial energy, low dimensionality and elastic distortions allow the growth of one-dimensional order, but inhibit the lateral translational symmetry of fully ordered 3D-crystals. Indeed, a reduction of the dimensionality of periodic ordering in solid phases could be a recurrent consequence of nanoconfinement for anisotropic molecules. For instance, the crystal structures of long alkane chains confined in Vycor glass retains the bulk crystalline closepacked structure in two directions only [10]. The splitting of the Bragg peak of the Ks phase confined in PSi revealed by X-rays scattering experiments is intriguing. The occurrence of two adjacent Bragg peaks cannot be easily assigned to a unique lamellar phase. One possible interpretation is that 8CB may be trapped in sensibly different configurations that are reminiscent of the Ks phase because of confinement-induced strains and most probably, quenched disorder effects. A second interpretation implies the K' phase. The high-*q* signatures of the K' phase have always be observed in co-existence with the lamellar Ks Bragg peak [18]. It suggests that the K' phase most likely nucleates from the Ks one. It could indeed correspond to the growth of a lateral ordering within the layers of the Ks phase, which does not alter the lamellar structure and preserve a Bragg reflection very close to 0.25Å$^{-1}$.

At variance to disordered porous materials, such as aerogels, the use of macroscopically aligned channels provides unique information on the molecular orientation within the lamellar structure of the Ks and K's phases. The planar spacing *d* of the K's and Ks phases are respectively 4.2 nm and 2.5 nm, enclosing the value for the SmA phase (3.1 nm). The structure of the SmA phase of 8CB corresponds to a smectic packing of partially overlapping couples of molecules, usually denoted as a $SmA_d$ type of smectic phase. For the K's phase, the planar spacing *d* of 4.2 nm corresponds to twice the length of the molecule and the crystallographic direction of the related Bragg peak is parallel to the pore axis. It is therefore consistent with a periodic smectic A-type ordering of layers with almost not overlapping molecules, i.e. a $SmA_2$ type of lamellar structure. Interestingly, the diffraction pattern of the Ks phase, corresponding to four symmetrically located peaks, is not consistent with a SmA type of lamellar structure but compares with the splitting observed for SmC phases [37]. If one assumes that during the SmA-Ks transition the nematic director is fixed by the unidirectional constraint of confinement, parallel to the pore axis and that the overlapping between molecules of adjacent layers is maintained, one may relate the tilt angle of the layers to the lattice constant as follows : $\cos^{1}(\frac{d_{Ks}}{d_{SmA}})$. These two assumptions lead to =41°, which is in quantitative agreement with the angular location ≈45° of the Bragg peaks measured by SANS in Fig. 12. It proves that the Ks phase is obtained in the nanochannel of PSi from the SmA phase by a simple shearing of the molecules along the nematic director that produces a tilt of the layer normal, while the director remains pinned along the pore axis direction. It also maintains the same molecular overlapping and period along the direction of the director as in the SmA phase. Interestingly, the opposite scenario has been reported for the SmA to SmC transition of $\bar{8}$S5 confined in anisotropic aerosils, which proceeds by a tilting of the director, while the layer normal is maintained [23]. This illustrates different energetic balances between random tilt fields and random positional fields in the cases of PSi and anisotropic aerosils.

## 5 Conclusions

Confinement of liquid crystals in unidirectional porous silicon mesoporous materials has recently deserved much interest [19,34]. Indeed, it has shown drastic changes of both the structure, dynamics and phase transitions in the 'high temperature range' where mesomorphic phases are encountered. This unusual behaviour has been related to the interplay of unidirectional confinement and quenched disorder effects. Such conditions have not been encountered is the previously studied systems, which are either homogeneous random porous materials [14,15] or unidirectional pores with weak disorder effects [12].

In the present study, we have studied the low temperature phase behaviour of the prototypical mesogenic system 8CB, in bulk and under spatial confinement in two types of uniaxial nanopores provided by porous silicon (PSi) and alumina anopore membranes. Both materials induce a strongly unidirectional confinement, which can be expressed in terms of finite size, surface interaction and low dimensionality effects. Additionally, the highly irregular inner surface of PSi introduces large quenched disorder effects, which are small for anopores.

We unambiguously show that unidirectional nanoconfinement promotes the formation of lamellar structures (phases K's and Ks), which are not normally encountered in bulk conditions. Bulk-like three-dimensional order can, however, be recovered in nanopores (K' and K phases). Indeed, the most metastable lamellar phases are preferentially obtained by confinement and by fast cooling rates, whereas slow cooling rates or heating after ageing at low temperature allow the formation of K' and K phases, the growth of which is kinetically controlled in confined geometry. Combining results obtained for different thermal treatments allows drawing a first tentative Gibbs diagram of the relative stability of these different phases.

The lamellar Ks phase can be obtained by quenching of the bulk or by confinement, which illustrates the apparent similarity between thermal quench and confinement effects. This is at variance with the infrequent K’s phase, which is formed in PSi only. The stability of the phase is weak and it eventually melts to the smectic phase on heating above 250K (i.e. more than 40 K below the melting point of the confined K phase). This suggests that stronger spatial confinement and quenched disorder effects introduced by PSi bring an additional constraint to full structural relaxation and extension of a completely ordered structure, therefore promoting the uncommon K’s phase. The specificity of PSi as compared to other unidirectional confined geometries has therefore some direct implications on both the smectic phase properties and the low temperature polymorphism [19,34].

The use of porous materials permeated by macroscopically aligned linear channels has provided invaluable information about the structure of these unusual lamellar phases, which goes beyond the present understanding of the rich polymorphism of 8CB in confinement, as essentially described in the case of random porous materials [17,18]. Ks and K’s phases present a preferential alignment of the nematic director along the pore axis. In addition, it proves that the Ks structure presents a smectic A type of arrangement, the interlayer spacing being consistent with almost no molecular overlapping between adjacent layers ($SmA_2$ type of lamellar structure). The K’s phase structural analysis suggests a tilted lamellar order with a tilt angle around 40°, which maintains the director orientation pinned in the direction of the pore axis and the interlayer molecular overlapping of the smectic A phase.

## Acknowledgements

We thank J. P. Ambroise and I. Mirebeau for the experiments performed at the Léon Brillouin Laboratory neutron source facility. Financial supports from the *Centre de Compétence C’Nano Nord-Ouest* and *Rennes Metropole* are expressly acknowledged. This study has been performed in the frame of a collaboration supported by a *Procope exchange program*.

## Table captions

**Table 1:** Compilation of the phase transition temperatures obtained in this study by calorimetry, neutron and X-rays scattering. Unless specified, the statistical experimental uncertainty is ±0.3K. In the case of PSi, phase A refers to the short range ordered smectic phase.

## Figures captions

**Fig. 1.** Thermograms of bulk 8CB measured by differential scanning calorimetry on cooling (dashed line) and on heating (full line) using a thermal rate of 2 $K.mn^{-1}$. inset : magnification of the high temperature region.

**Fig. 2.** Transition temperatures of 8CB measured by differential scanning calorimetry at different rates during cooling (open symbols) and during heating (filled symbols). (a) low temperature region : crystal K-smectic transition (triangle) of bulk 8CB (solid lines). Crystals K,K'-smectic transitions (triangle) and crystal Ks-smectic transition of 8CB confined in porous alumina anopore (dashed lines). (b) high temperature region : nematic-isotropic transition (squares) and smectic-nematic transition (circles) of bulk 8CB (solid lines) and 8CB confined in porous alumina anopore (dashed lines).

**Fig. 3.** Small angle neutron scattering patterns of (a) bulk 8CB at 315 K (isotropic phase) (b) bulk 8CB at 300K (smectic phase), (c) bulk 8CB at 300K under application of a magnetic field of 0.6 T (aligned smectic monodomain).

**Fig. 4.** Double-axis neutron scattering patterns of bulk 8CB at different temperatures on heating (mean heating rate 0.7 K. $min^{-1}$).

**Fig. 5.** Thermograms of 8CB confined in porous alumina anopores measured by differential scanning calorimetry on cooling (dashed line) and on heating (solid line with striped area) using a thermal rate of 10 $K.mn^{-1}$. The second thermogram on heating (solid line with grey shaded area) is acquired after a prior heating and ageing at 288K. Inset: same thermograms measured on heating without (solid line with striped area) and with (solid line with grey shaded area) a prior thermal treatment at 288K and using a thermal rate of 2 $K.mn^{-1}$.

**Fig. 6.** Small angle neutron scattering patterns of 8CB confined into porous alumina anopores at 300K (smectic phase). The sharpest Bragg peak reflects the full alignment of the director along the unidirectional pore axis.

**Fig. 7.** Double-axis neutron scattering patterns of 8CB confined in alumina anopores at different temperatures on heating (mean heating rate 0.7 K. $min^{-1}$). (a) in normal incidence, (b) in grazing incidence. Inset: magnification of the low-*q* region.

**Fig. 8.** Temperature variation of the maximum intensity of a selection of characteristic Bragg peaks at $q$=0.2$Å^{-1}$ for the smectic (circle), at $q$=0.48$Å^{-1}$ for the K (triangle), at $q$=0.65$Å^{-1}$ for the K’ (diamond) and at $q$=0.25Å$^{-1}$ for the Ks (square) phases of bulk 8CB (filled symbols) and 8CB confined in anopores (open symbols) during heating. The intensities have been extracted from the double-axis neutron scattering patterns displayed in Figs. 4 and 7. Each temperature dependent Bragg peak intensity is rescaled with respect to its largest value (for clarity the peak at $q$=0.25Å$^{-1}$ is not scaled to unity but to 0.5).

**Fig. 9.** Small angle neutron scattering patterns of 8CB confined in porous silicon at different temperature on cooling from 340K to 180K. The occurrence of Bragg peaks during cooling is highlighted by dashed lines at 290K (Short range smectic phase), 240K (Ks phase) and 220K (K’s phase).

**Fig. 10.** Temperature variation of the maximum intensity of a selection of characteristic Bragg peaks of the smectic ($q$=0.2Å$^{-1}$), K’s ($q$=0.15Å$^{-1}$) and Ks ($q$=0.25Å$^{-1}$) phases of 8CB confined in porous silicon (filled symbols) during cooling. The smectic region of bulk 8CB is added for comparison (open diamonds). The intensities have been extracted from double-axis neutron scattering patterns. Each smectic Bragg peak intensity is rescaled with respect to its largest value, whereas an additional scaling factor of 0.85 (resp. 0.15) for the K’s (resp. Ks) Bragg peak intensity is used to illustrate the fraction of the system converted into these two phases.

**Fig. 11.** Temperature variation of the maximum intensity of a selection of characteristic Bragg peaks of 8CB confined in porous silicon during cooling (filled circles) and subsequent heating (open circle). The intensities have been extracted from double-axis neutron scattering. Each smectic Bragg peak intensity is rescaled with respect to its largest value, whereas an additional scaling factor of 0.85 (resp. 0.15) for the K’s (resp. Ks) Bragg peak intensity is used to illustrate the fraction of the system converted into these two phases. (a) Intensity at $q$=0.2Å$^{-1}$ associated to the smectic phase Bragg peak. The smectic region of bulk 8CB is added for comparison (open diamonds) ; (b) Intensity at $q$=0.15Å$^{-1}$ associated to the K’s phase Bragg peak. The crystal K region of bulk 8CB is added for comparison (open diamonds); (c) Intensity at $q$=0.25Å$^{-1}$ (triangles up) and at $q$=0.48Å$^{-1}$ (triangles down) associated to the Ks phase and K phase Bragg peaks respectively.

**Fig. 12.** X-rays diffraction spectra of 8CB confined in porous silicon at 280K (upper curve), at 245 K on cooling (middle curve), and at 280K after re-heating (lower curve), with a mean thermal rate of about 0.01 K.min$^{-1}$.

**Fig. 13.** Schematic Gibbs energy diagram showing the relative stability of the different phases of 8CB confined in unidirectional nanopores, combining results acquired for 8CB confined in alumina anopores and in porous silicon.

| | | | | | | | | | |
|---|---|---|---|---|---|---|---|---|---|
| | | | | | | | **K-A** | **A-N** | **N-I** |
| **BULK** | | | | | | | | | |
| **(**from ref. [25]) | | | | | | | 295.2 | 307.0 | 314.0 |
| (from ref [29]) | | | | | | | 293.0 | 305.6 | 313.0 |
| **BULK** | Heating (DSC, Tmax) 10K.min$^{-1}$ | | | | | | 296.55 | 307.65 | 315.15 |
| (This study) | Heating (DSC, Tmax) extrap. to 0K.min$^{-1}$ | | | | | | 294.05 | 306.25 | 313.4 |
| | Cooling (DSC, Tmax) extrap. to 0K.min$^{-1}$ | | | | | | 268.15 | 306.25 | 313.25 |
| | Heating (Neutrons, $q$=0.2Å$^{-1}$) | | | | | | 294.8 | | |
| | Heating (Neutrons, $q$=0.48Å$^{-1}$) | | | | | | 294.1 | 306.0 | - |
| | | | **Ks-A** | | | **K'-A** | **K-A** | **A-N** | **N-I** |
| **Anopore** | Heating (DSC, Tmax) 10K.min$^{-1}$ | | 286.1 | | | 291.7 | 292.5 | 304.9 | 311.7 |
| | Heating (DSC, Tmax) extrap. to 0K.min$^{-1}$ | | 284.7 | | | 290.7 | 291.4 | - | - |
| | Heating (Neutrons, $q$=0.2Å$^{-1}$) | | | | | | 293.5 | 306.0 | |
| | Heating (Neutrons, $q$=0.48Å$^{-1}$) | | | | | | 293.2 | | |
| | Heating (Neutrons, $q$=0.65Å$^{-1}$) | | | | | 293.5 | | | |
| | Heating (Neutrons, $q$=0.25Å$^{-1}$) | | 287±1 | | | | | | |
| | | **K's-A** | | | | | | **A-N** | |
| **Porous** | Cooling (Neutrons, $q$=0.2Å$^{-1}$) | 231 ±1 | | | | | | 260-310 | |
| **Silicon** | Cooling (Neutrons, $q$=0.15Å$^{-1}$) | 232±1 | | | | | | | |
| (85 % of the | Heating (Neutrons, $q$=0.2Å$^{-1}$) | 255 ±1 | | | | | | 260-310 | |
| confined | Heating (Neutrons, $q$=0.15Å$^{-1}$) | 251 ±1 | | | | | | | |
| phase) | | | | | | | | | |
| | | | **Ks-A** | **Ks-K** | **K'-K** | **K'-A** | **K-A** | **A-N** | |
| **Porous** | Cooling (Neutrons, $q$=0.25Å$^{-1}$, $q$=0.2Å$^{-1}$) | | 250±5 | | | | | 260-310 | |
| **Silicon** | | | | | | | | | |
| (15 % of the | Heating (Neutrons, $q$=0.25Å$^{-1}$, $q$=0.48Å$^{-1}$) | | | 280±5 | | | 296±5 | | |
| confined | | | | | | | | | |
| phase) | | | | | | | | | |
| **Porous** | Cooling ($q$=0.25Å$^{-1}$) | | 272±3 | | | | | | |
| **Silicon** | Cooling ($q$=1.29Å$^{-1}$) | | | | | 269±1 | | | |
| Ultra-slow | Cooling ($q$=1.62Å$^{-1}$) | | | | | 271±5 | | | |
| rates by X-rays | Heating ($q$=0.25Å$^{-1}$,$q$=0.48Å$^{-1}$) | | | 270±3 | | | | | |
| | Heating ($q$=1.29Å$^{-1}$) | | | | 270±2 | | | | |
| | Heating ($q$=1.62Å$^{-1}$) | | | | 266±3 | | | | |
| | Heating ($q$=0.48Å$^{-1}$) | | | | | | 295±2 | | |

## Table 1

**Rich polymorphism of 8CB confined in two types of unidirectional nanopores** R. Guégan et al.

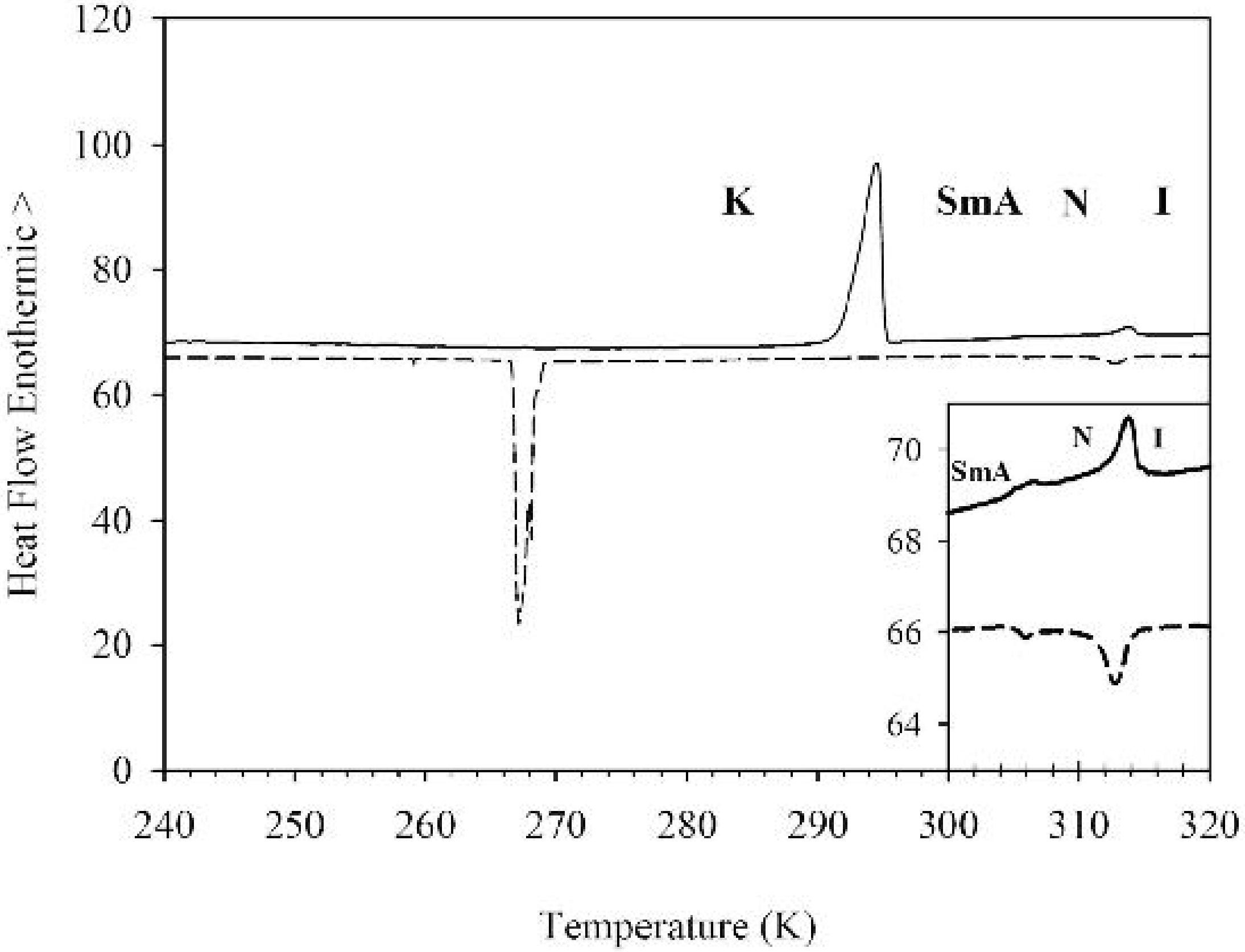

Figure 1
**Rich polymorphism of 8CB confined in two types of unidirectional nanopores** R. Guégan et al.

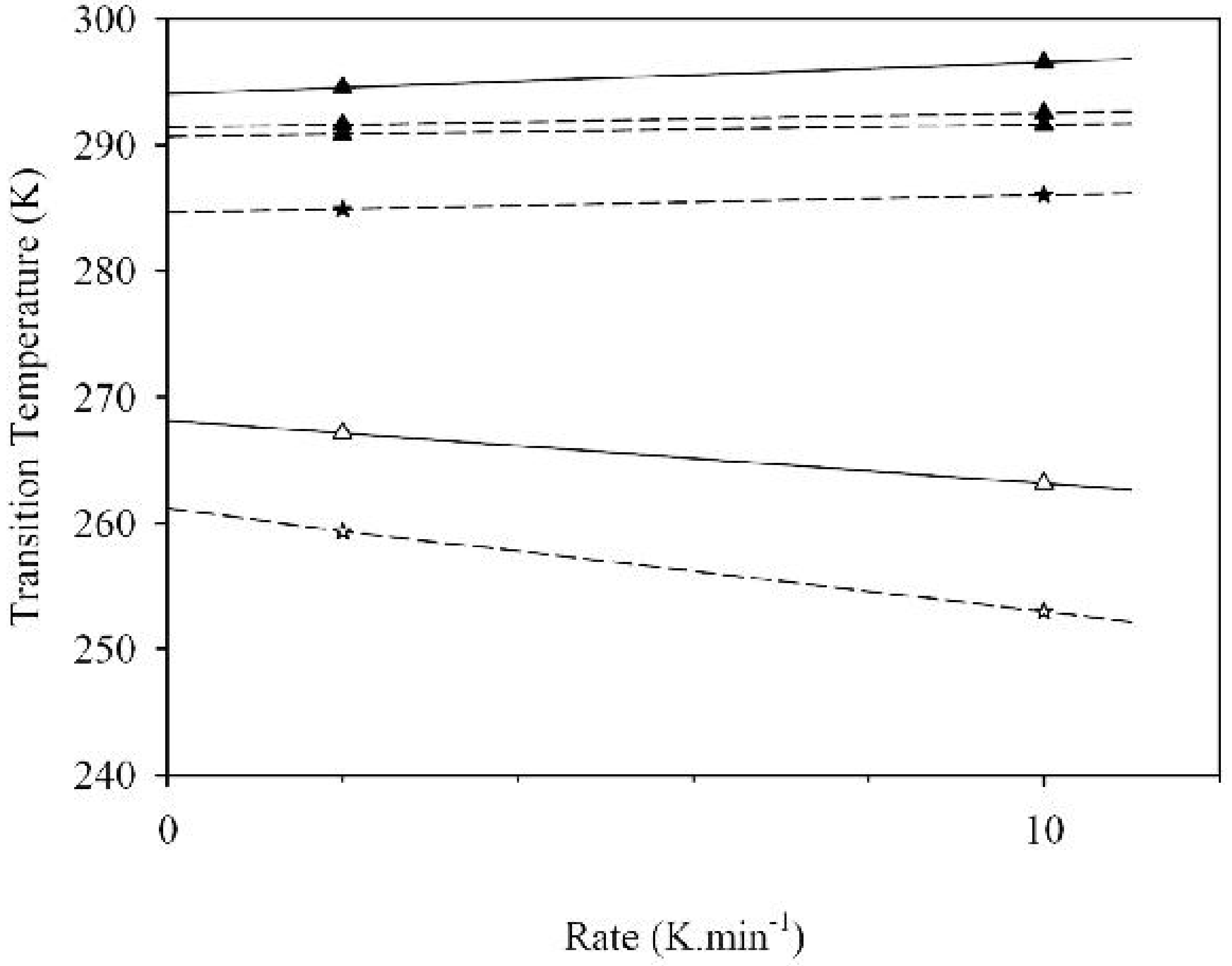

Figure 2,a

**Rich polymorphism of 8CB confined in two types of unidirectional nanopores** R. Guégan et al.

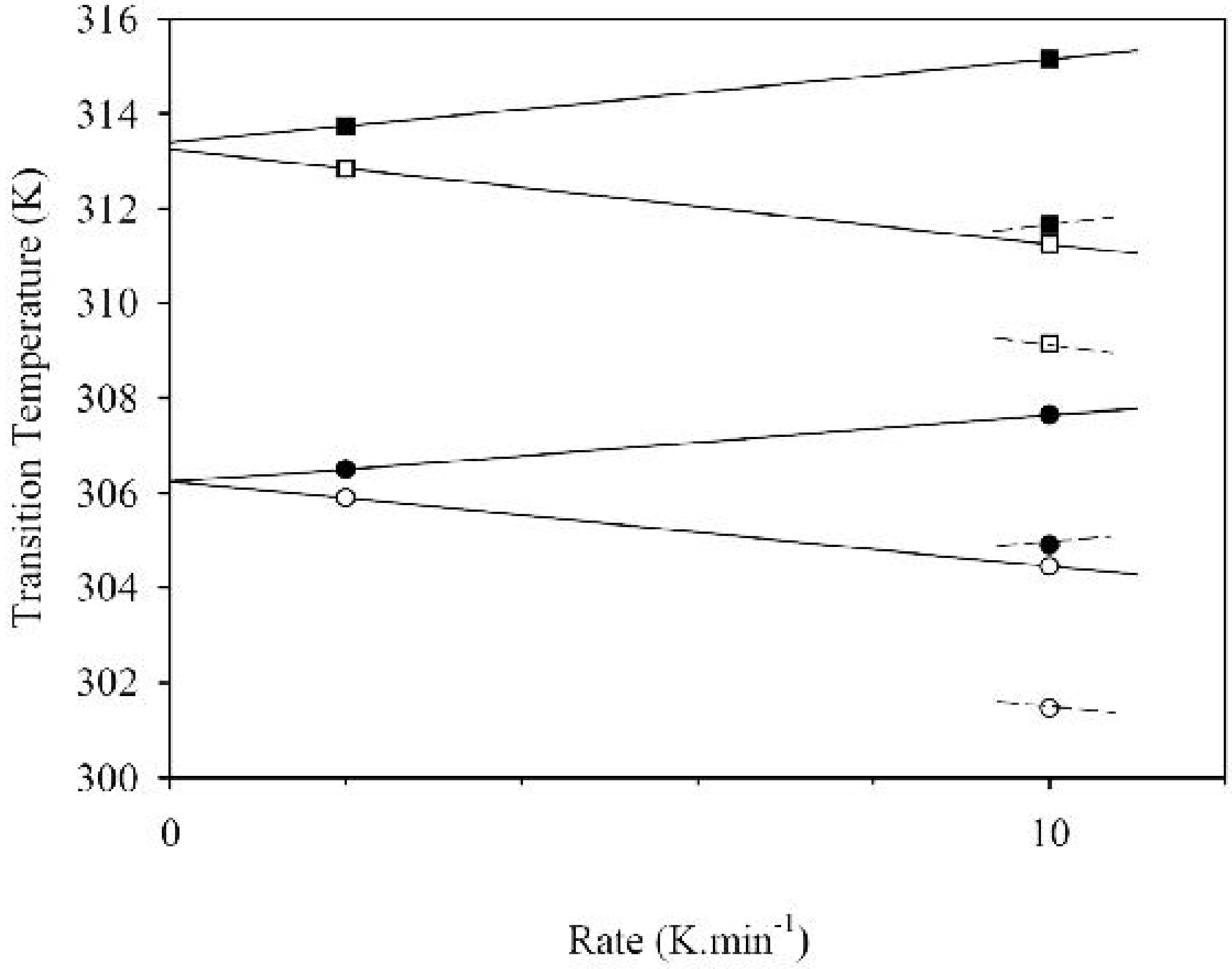

Figure 2,b

**Rich polymorphism of 8CB confined in two types of unidirectional nanopores** R. Guégan et al.

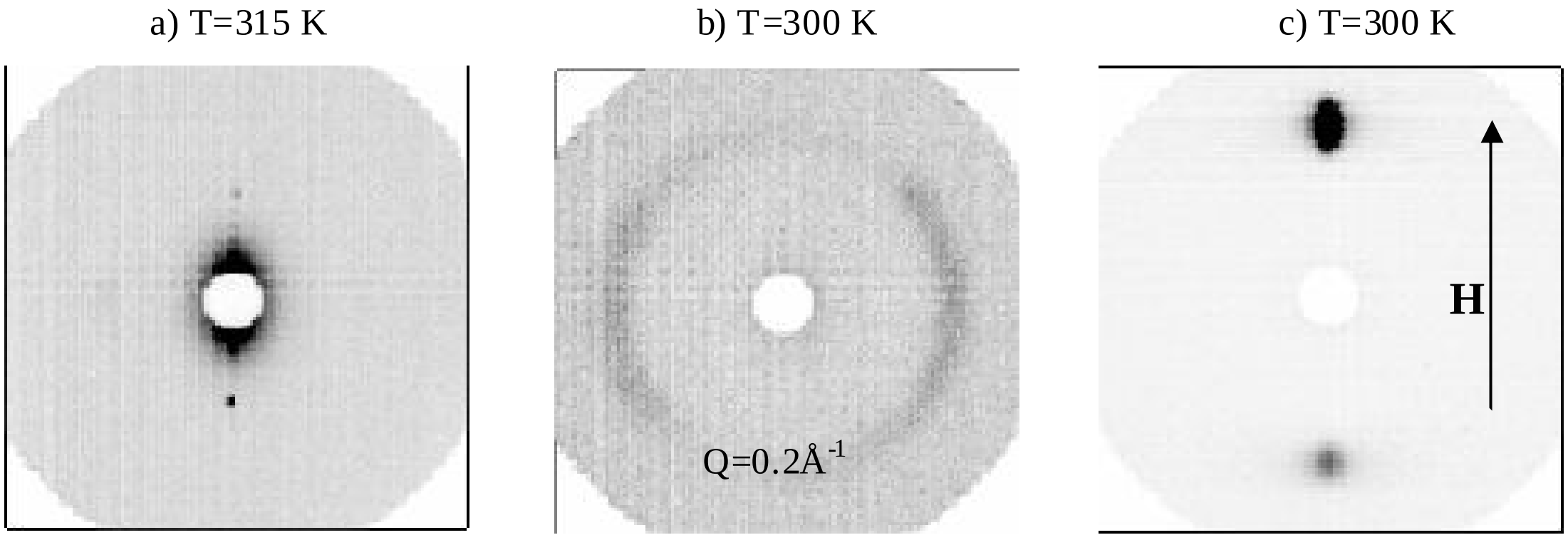

Figure 3
**Rich polymorphism of 8CB confined in two types of unidirectional nanopores**
R. Guégan et al.

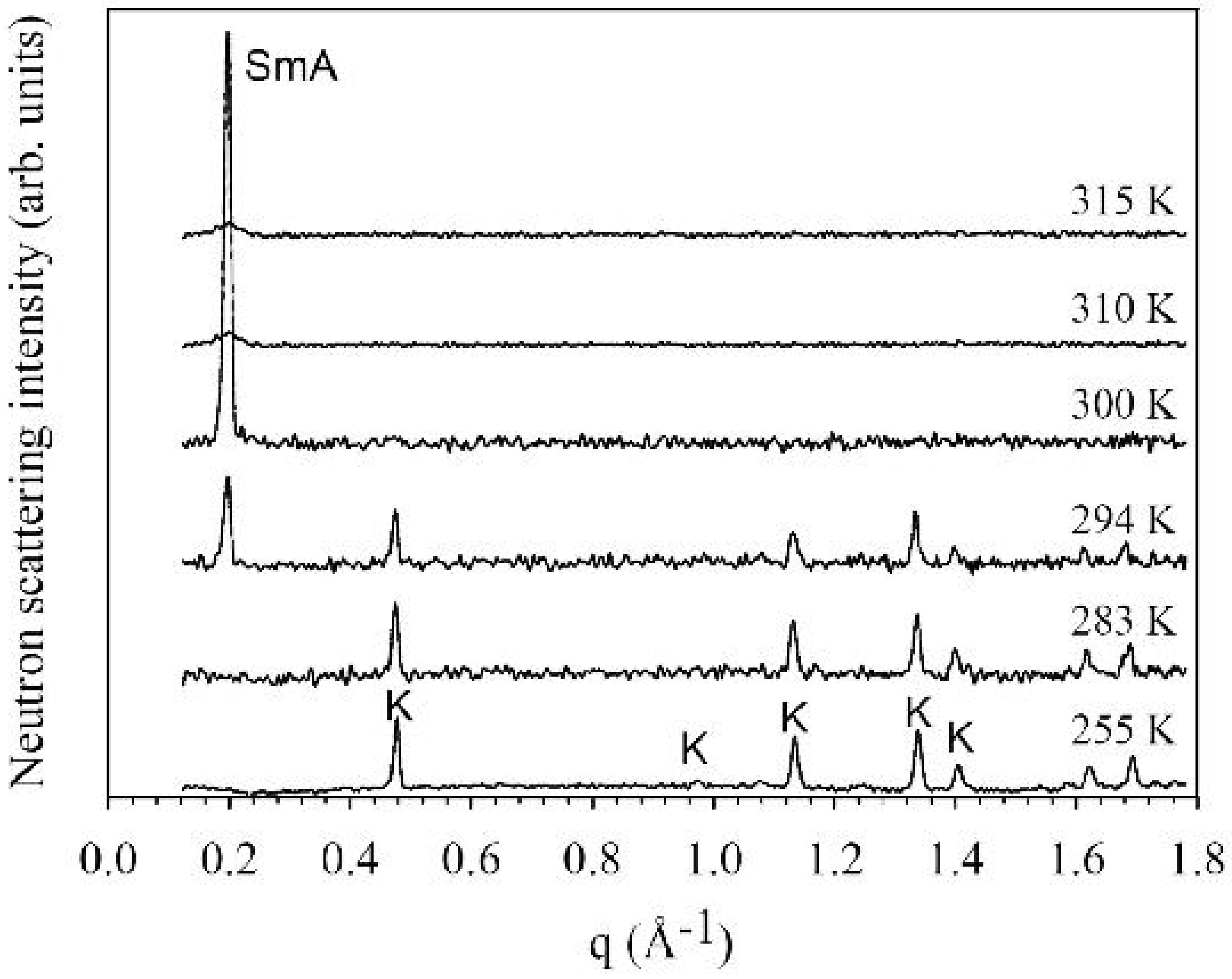

Figure 4

**Rich polymorphism of 8CB confined in two types of unidirectional nanopores**

R. Guégan et al.

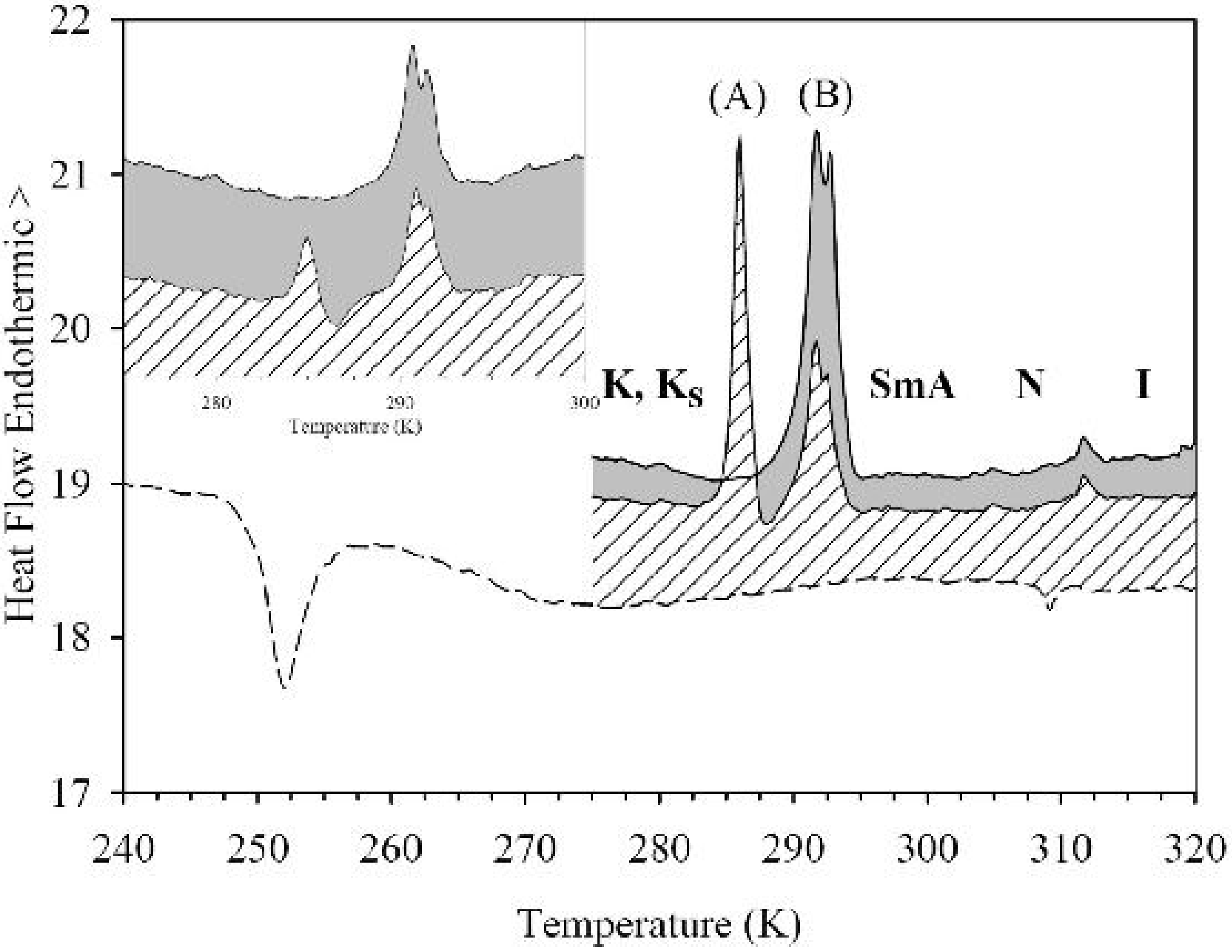

Figure 5

**Rich polymorphism of 8CB confined in two types of unidirectional nanopores**

R. Guégan et al.

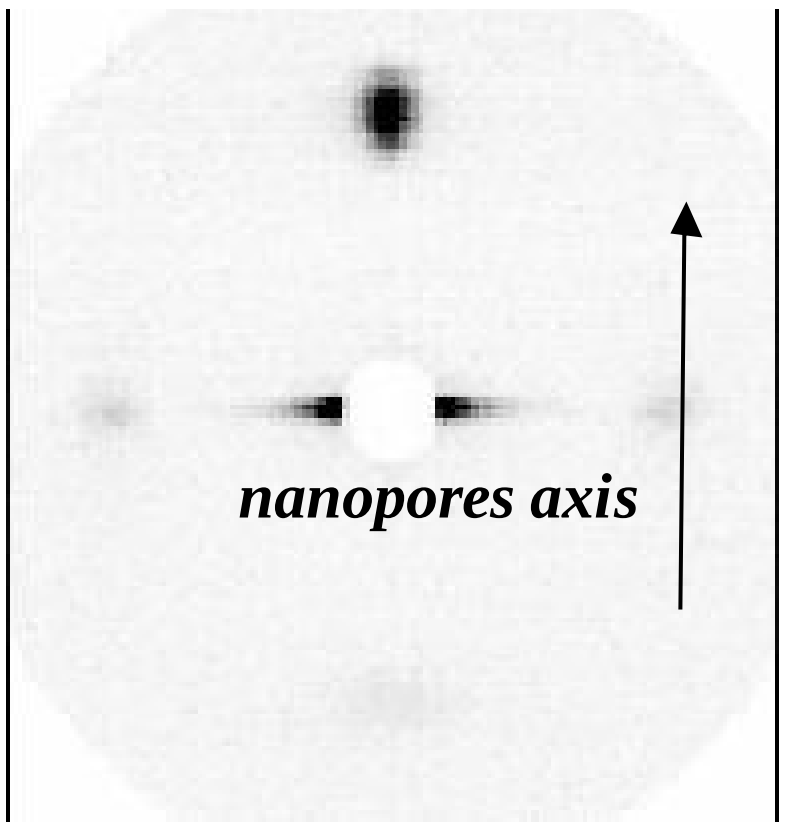

Figure 6
**Rich polymorphism of 8CB confined in two types of unidirectional nanopores**
R. Guégan et al.

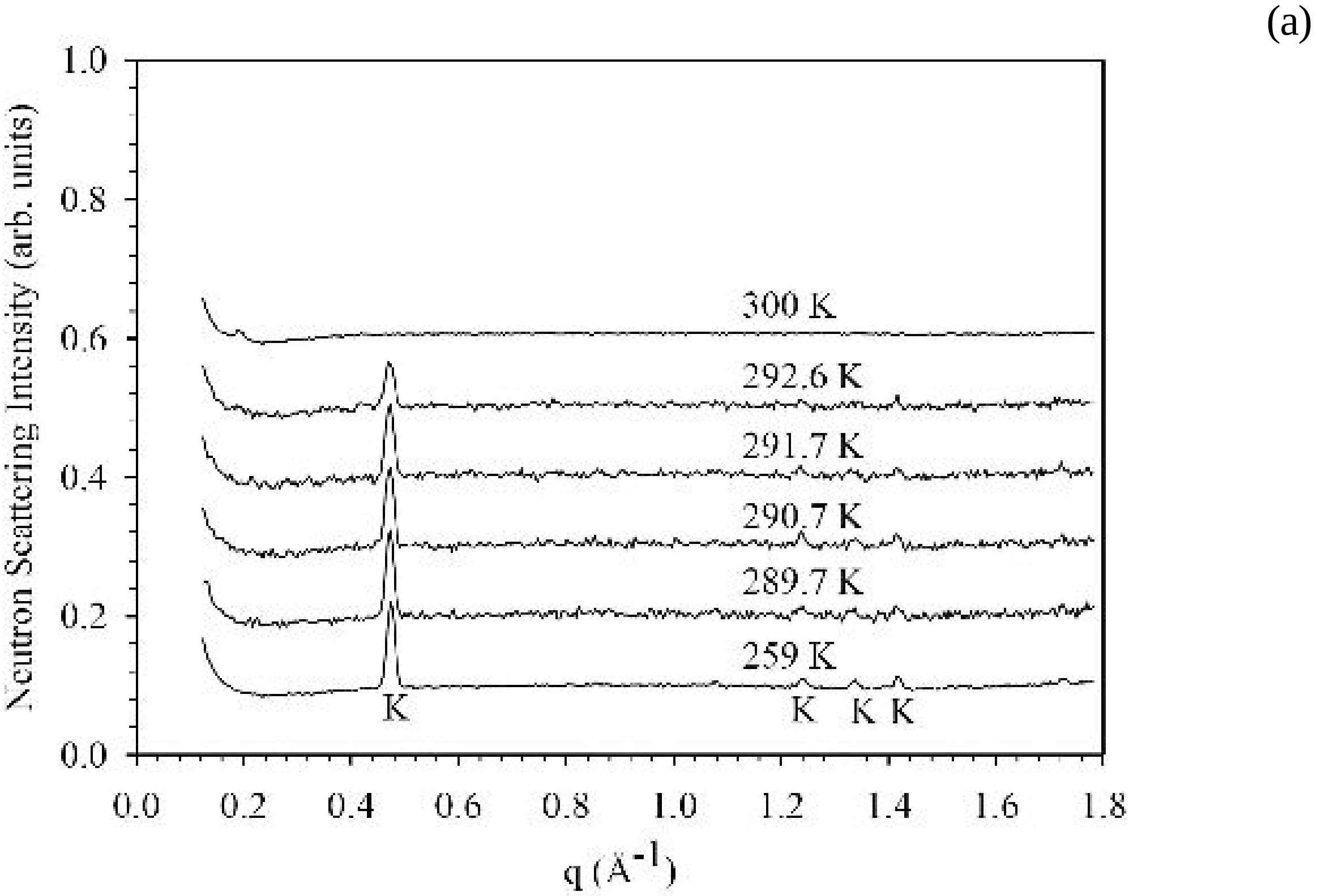

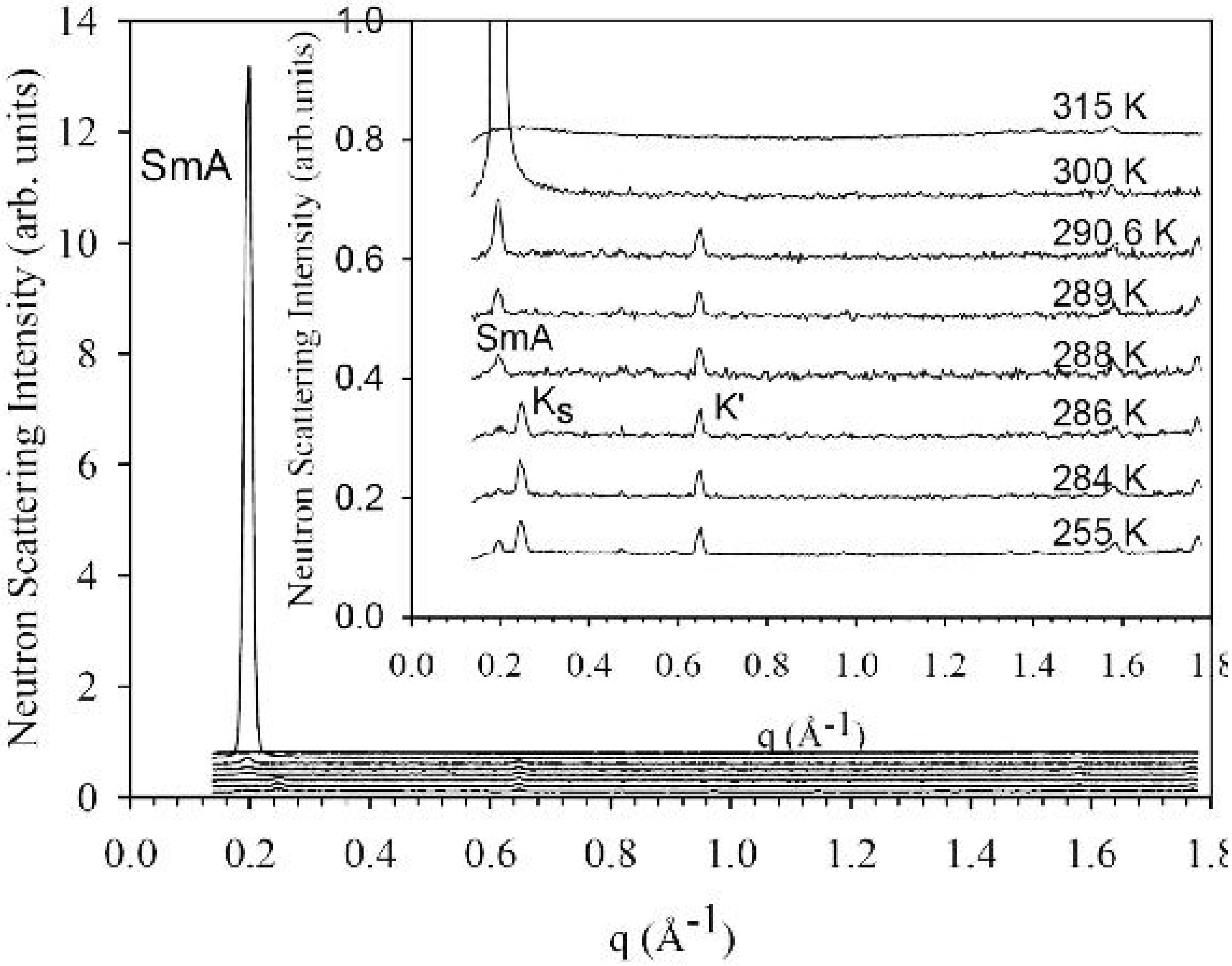

(b)

Figure 7

**Rich polymorphism of 8CB confined in two types of unidirectional nanopores**

R. Guégan et al.

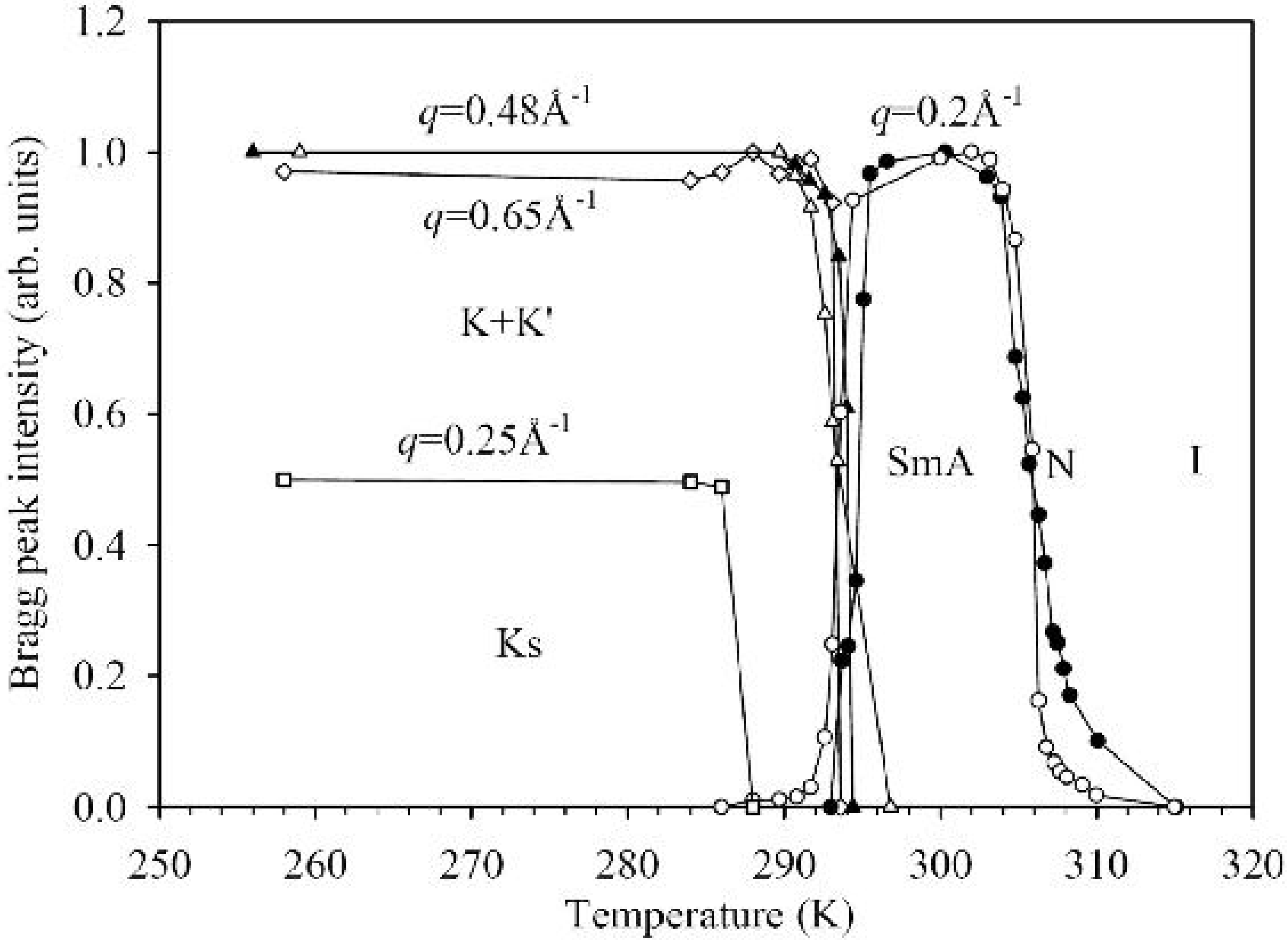

Figure 8
**Rich polymorphism of 8CB confined in two types of unidirectional nanopores**
R. Guégan et al

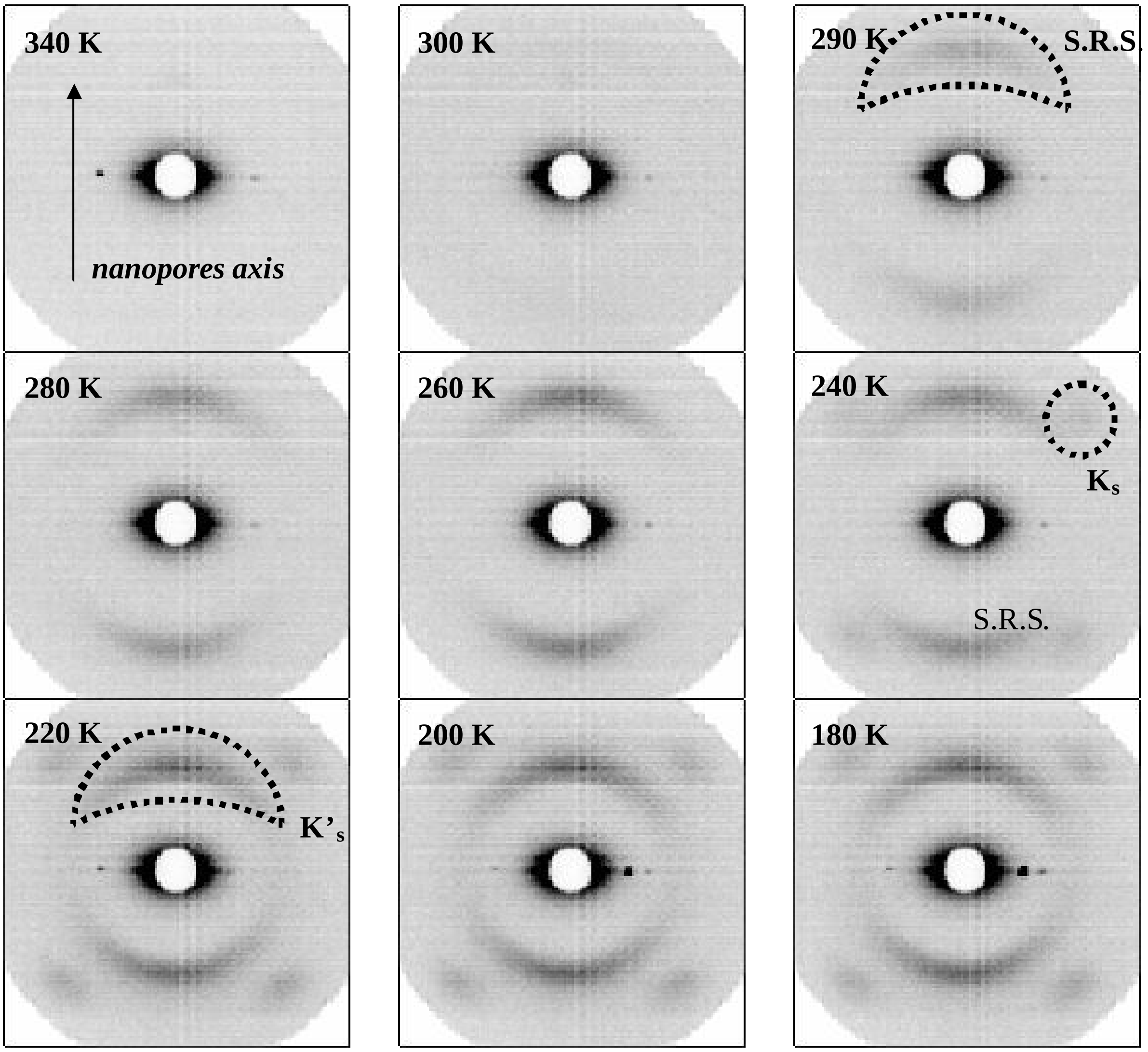

Figure 9
**Rich polymorphism of 8CB confined in two types of unidirectional nanopores**
R. Guégan et al.

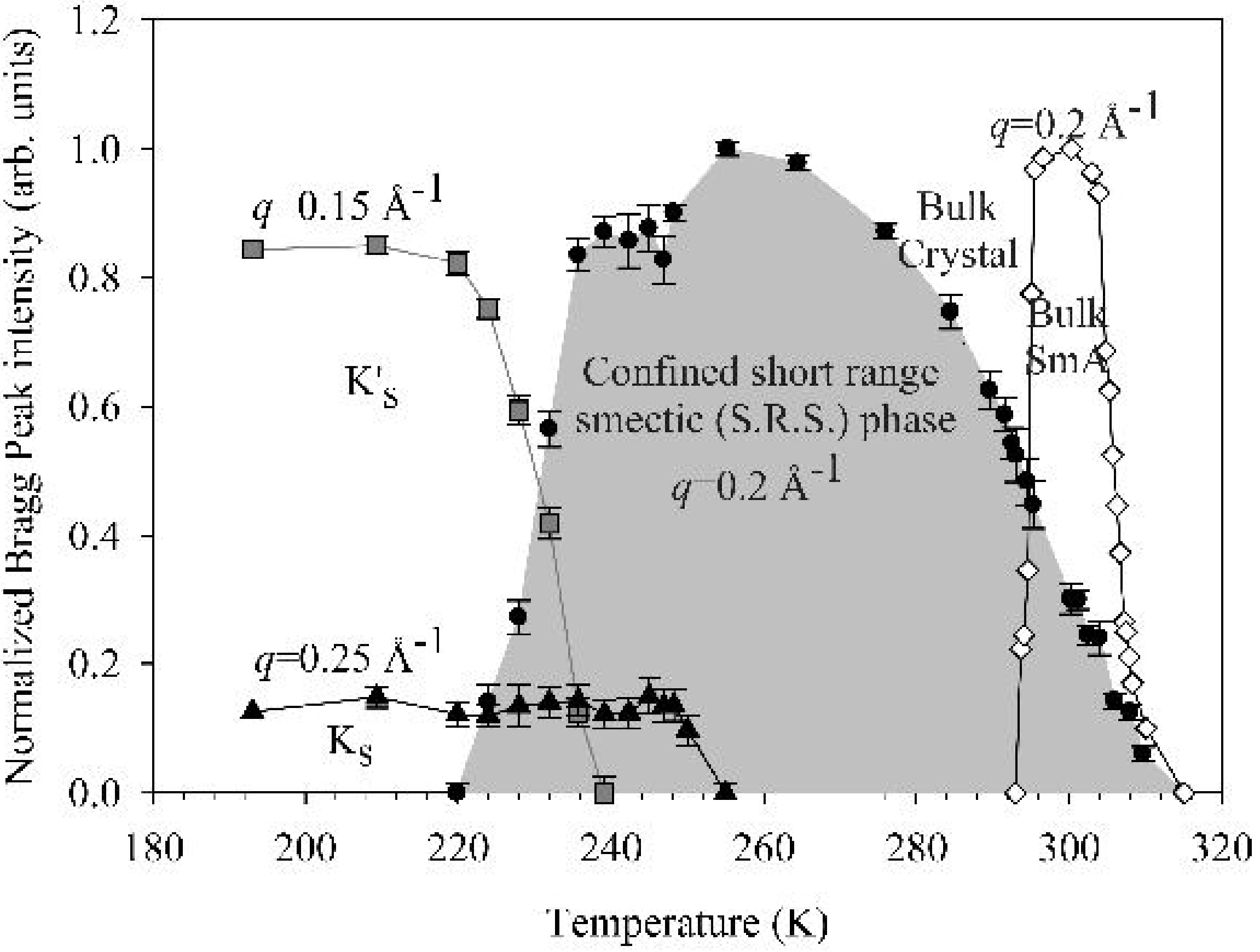

Figure 10

**Rich polymorphism of 8CB confined in two types of unidirectional nanopores**

R. Guégan et al.

(a)

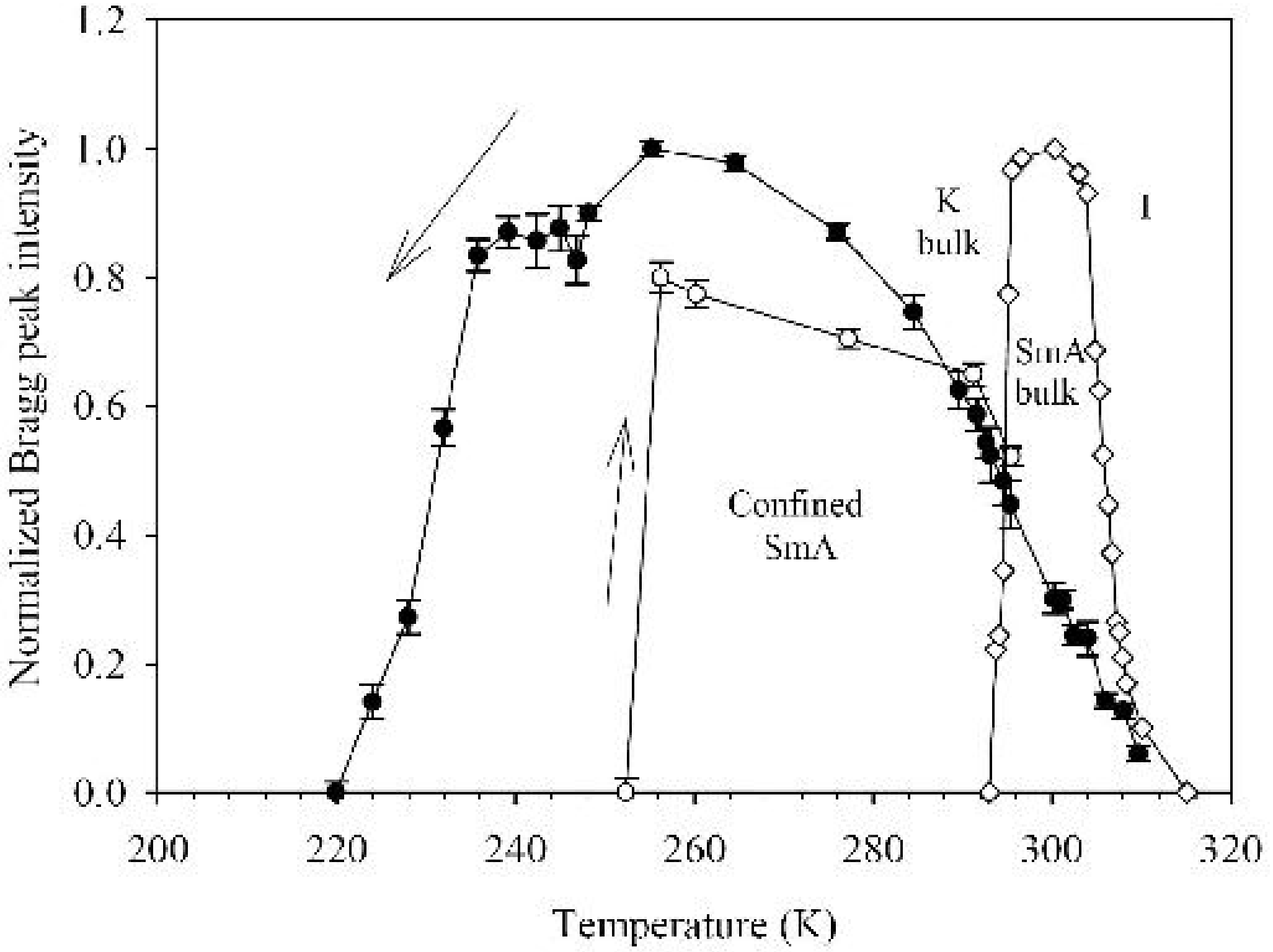

(b)

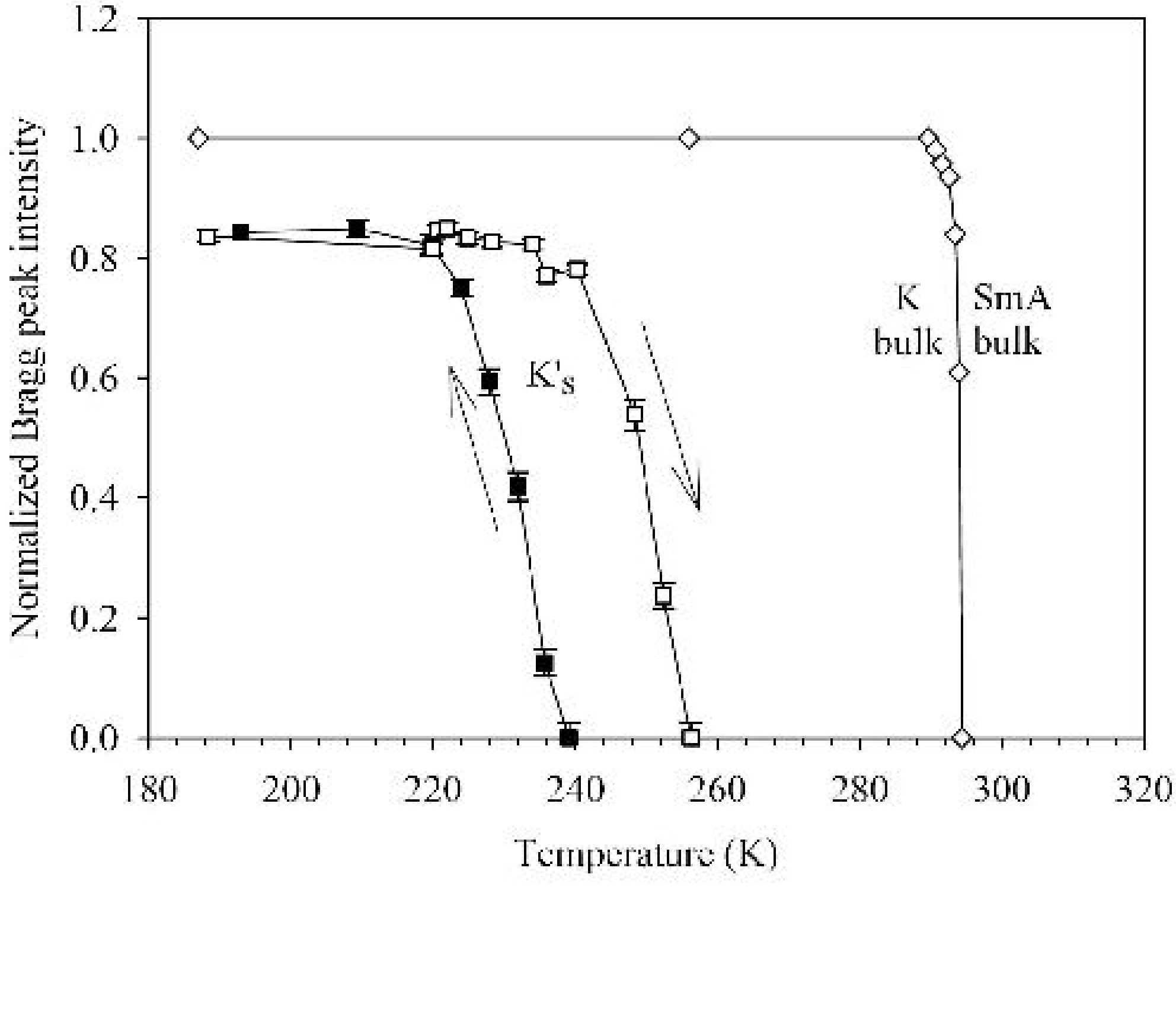

(c)

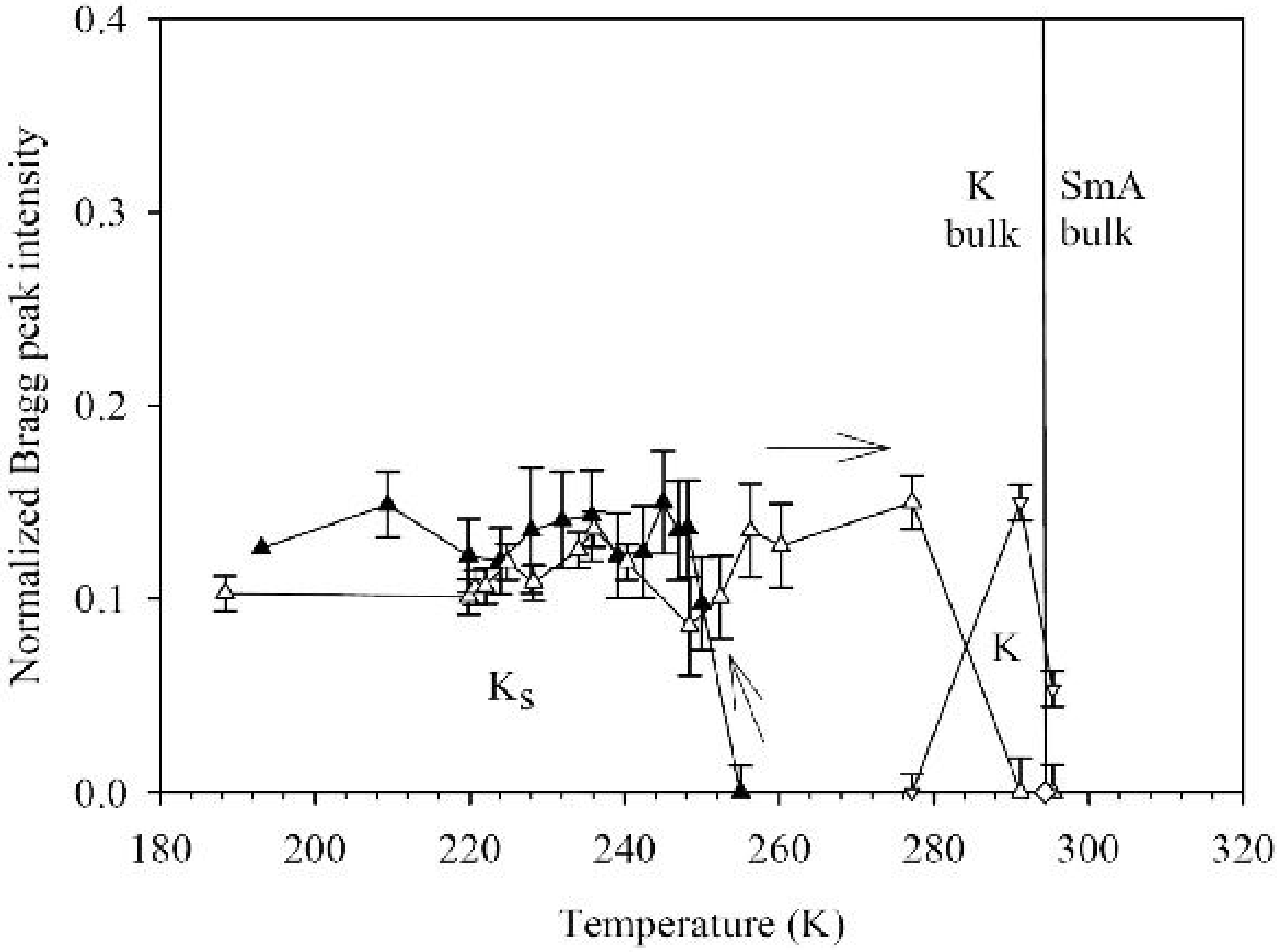

Figure 11
**Rich polymorphism of 8CB confined in two types of unidirectional nanopores**
R. Guégan et al.

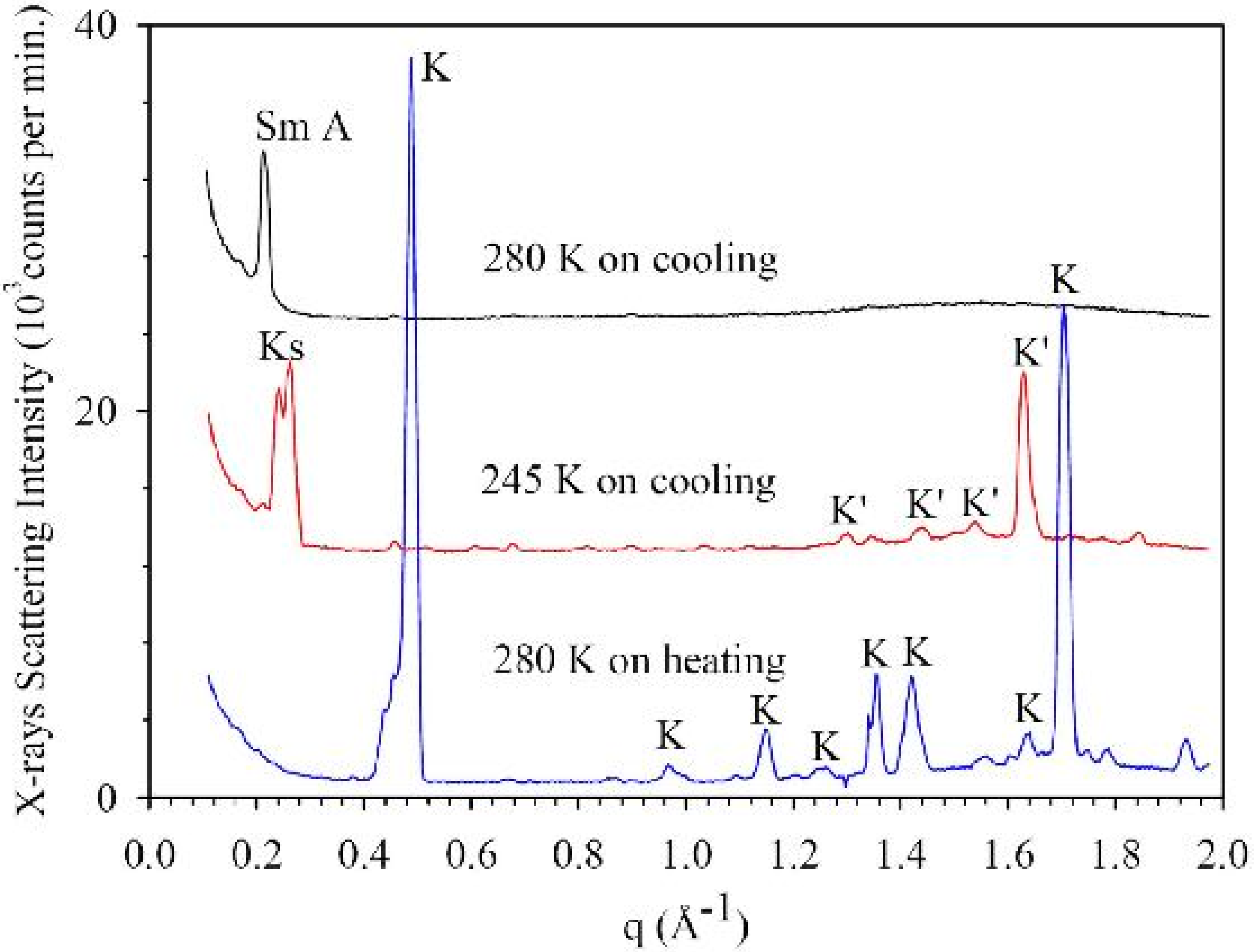

Figure 12
**Rich polymorphism of 8CB confined in two types of unidirectional nanopores**
R. Guégan et al.

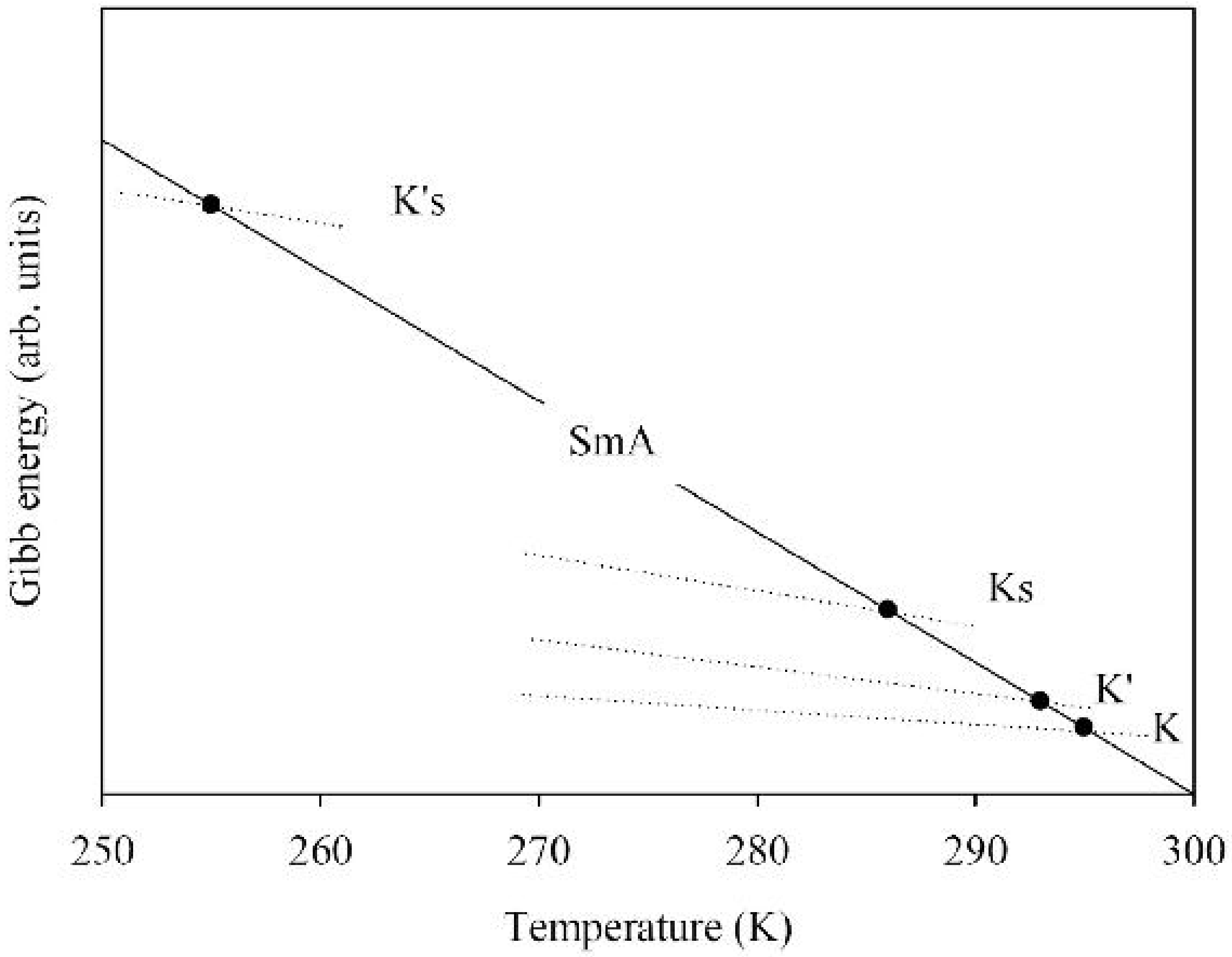

Figure 13

**Rich polymorphism of 8CB confined in two types of unidirectional nanopores**

R. Guégan et al.